\documentclass[preprint]{aastex63}
\usepackage{graphicx}
\usepackage{psfig}
\usepackage{natbib}
\usepackage{subfigure} 

\newcommand{\gsim}{\mbox{\hspace{.2em}\raisebox{.5ex}{$>$}\hspace{-.8em}\raisebox{-.5ex}{$\sim$}\hspace{.2em}}}
\newcommand{\lsim}{\mbox{\hspace{.2em}\raisebox{.5ex}{$<$}\hspace{-.8em}\raisebox{-.5ex}{$\sim$}\hspace{.2em}}}

\newcommand{\E}[1]{\times 10^{#1}}
\newcommand{\twCO}{$^{12}$CO}  \newcommand{\thCO}{$^{13}$CO}

\newcommand{\HI}{\mbox{H\,\textsc{i}}}

\newcommand{\s}{\,{\rm s}}      \newcommand{\ps}{\,{\rm s}^{-1}}
    \newcommand{\Msun}{M_{\odot}}   
    \newcommand{\km}{\,{\rm km}}

        \newcommand{\K}{\,{\rm K}}
    
    \newcommand{\G}{\,{\rm G}}

\begin{document}

\title{
CO Emission Delineating the Interface between the
Milky Way Nuclear Wind Cavity and the Gaseous Disk
}

\shorttitle{Molecular walls of the Galactic-Center superbubbles}

\correspondingauthor{Yang Su}
\email{yangsu@pmo.ac.cn}

\author[0000-0002-0197-470X]{Yang Su}
\affil{Purple Mountain Observatory and Key Laboratory of Radio Astronomy,
Chinese Academy of Sciences, 10 Yuanhua Road, Nanjing 210023, China}

\author{Shiyu Zhang}
\affiliation{Purple Mountain Observatory and Key Laboratory of Radio Astronomy,
Chinese Academy of Sciences, 10 Yuanhua Road, Nanjing 210023, China}
\affiliation{School of Astronomy and Space Science, University of Science and
Technology of China, 96 Jinzhai Road, Hefei 230026, China}

\author{Ji Yang}
\affiliation{Purple Mountain Observatory and Key Laboratory of Radio Astronomy,
Chinese Academy of Sciences, 10 Yuanhua Road, Nanjing 210023, China}
\affiliation{School of Astronomy and Space Science, University of Science and
Technology of China, 96 Jinzhai Road, Hefei 230026, China}

\author{Qing-Zeng Yan}
\affiliation{Purple Mountain Observatory and Key Laboratory of Radio Astronomy,
Chinese Academy of Sciences, 10 Yuanhua Road, Nanjing 210023, China}

\author{Yan Sun}
\affiliation{Purple Mountain Observatory and Key Laboratory of Radio Astronomy,
Chinese Academy of Sciences, 10 Yuanhua Road, Nanjing 210023, China}

\author{Hongchi Wang}
\affiliation{Purple Mountain Observatory and Key Laboratory of Radio Astronomy,
Chinese Academy of Sciences, 10 Yuanhua Road, Nanjing 210023, China}
\affiliation{School of Astronomy and Space Science, University of Science and
Technology of China, 96 Jinzhai Road, Hefei 230026, China}

\author{Shaobo Zhang}
\affiliation{Purple Mountain Observatory and Key Laboratory of Radio Astronomy,
Chinese Academy of Sciences, 10 Yuanhua Road, Nanjing 210023, China}

\author{Xuepeng Chen}
\affiliation{Purple Mountain Observatory and Key Laboratory of Radio Astronomy,
Chinese Academy of Sciences, 10 Yuanhua Road, Nanjing 210023, China}
\affiliation{School of Astronomy and Space Science, University of Science and
Technology of China, 96 Jinzhai Road, Hefei 230026, China}

\author{Zhiwei Chen}
\affiliation{Purple Mountain Observatory and Key Laboratory of Radio Astronomy,
Chinese Academy of Sciences, 10 Yuanhua Road, Nanjing 210023, China}

\author{Xin Zhou}
\affiliation{Purple Mountain Observatory and Key Laboratory of Radio Astronomy,
Chinese Academy of Sciences, 10 Yuanhua Road, Nanjing 210023, China}

\author{Lixia Yuan}
\affiliation{Purple Mountain Observatory and Key Laboratory of Radio Astronomy,
Chinese Academy of Sciences, 10 Yuanhua Road, Nanjing 210023, China}

\begin{abstract}
Based on the MWISP survey, we study high-$z$ CO emission toward the tangent points, 
in which the distances of the molecular clouds (MCs) are well determined.
In the region of $l=12^{\circ}$--$26^{\circ}$ and $|b|\lsim5\fdg1$,
a total of 321 MCs with $|z|\gsim$110~pc are identified, of which nearly 30 extreme
high-$z$ MCs (EHMCs at $|z|\gsim$260~pc) are concentrated in 
a narrow region of $R_{\rm GC}\sim$2.6--3.1~kpc.
The EHMC concentrations, together with other high-$z$ MCs at $R_{\rm GC}\lsim$2.3--2.6~kpc,
constitute molecular crater-wall structures surrounding the edges of the \HI\ voids 
that are physically associated with the Fermi bubbles.
Intriguingly, some large high-$z$ MCs, which lie in the crater walls 
above and below the Galactic plane, show cometary structures with the head 
toward the plane, favouring the scenario that the entrained molecular gas 
moves with the multiphase flows from the plane to the high-$z$ regions.
We suggest that the Milky Way nuclear wind has 
a significant impact on the Galactic gaseous disk.
The powerful nuclear wind at $\sim$3--6~Myr ago is likely responsible for
the observational features:  
(1) the enhanced CO gas lying in the edges of the \HI\ voids, 
(2) the deficiency of atomic and molecular gas within $R_{\rm GC}\lsim$3~kpc, 
(3) the possible connection between the EHMC concentrations and the 3-kpc arm, 
and (4) the elongated high-$z$ MCs with the tail pointing away from the Galactic plane.

\end{abstract}


\section{Introduction}
Galactic nuclear outflows and winds, which are powered by energetic processes 
in the central regions of galaxies, 
have dominated the transfer of mass, energy,
momentum, and metals from the disk to the halo and even the
intergalactic medium (IGM). The feedback from the nuclear activity is also crucial for 
the formation and evolution of galaxies and the IGM 
\citep[e.g.,][]{2005ARA&A..43..769V,2008MNRAS.387..639H,2018MNRAS.474.3673R,2021AN....342.1135M}. 

The Galactic center (GC) is the nearest laboratory to study the details 
of the feedback effects from the nuclear region of a galaxy.
Over the past several decades, a large number of exciting findings 
on the multiphase outflows from the GC have been revealed based on the compelling 
multiwavelength observations of the ground-based telescopes 
\citep[e.g.,][]{1984Natur.310..568S,2013Natur.493...66C,2016ApJ...831...72H,
2019Natur.573..235H,2020ApJ...899L..11K} and the space observatories
\citep[e.g.,][]{2000ApJ...540..224S,2003ApJ...582..246B,2006ApJ...646..951K,
2010ApJ...724.1044S,2013ApJ...779...57K,2014ApJ...793...64A,2015ApJ...799L...7F,
2017ApJ...834..191B,2020Natur.588..227P}.
The various studies show that past energetic events 
(e.g., AGN-driven and/or starburst-driven winds) at about several to tens of Myr ago 
may be responsible for the nuclear outflow/wind phenomena and 
the related large-scale structures \citep[see, e.g., the recent reviews by][]
{2020A&ARv..28....2V,2021A&A...646A..66P}.

The above studies mainly focus on the radio continuum, IR dust, optical, 
X-ray, $\gamma$-ray emission, and some UV absorption lines on 
the outflow structures of the Milky Way. 
On the other hand, the traditional tracers of the neutral atomic and molecular gas 
(i.e., 21~cm \HI\ and 2.6~mm CO emission) should be also very useful for revealing 
the spatial and dynamical features of the Galactic nuclear wind on a large scale.
In fact, the absence of high-$z$ atomic gas within the inner disk 
(i.e., Galactocentric distance of $R_{\rm GC}\lsim$~3~kpc) has been found 
by \citet{1984ApJ...283...90L} according to the atomic gas distribution.
The molecular gas in the inner 3~kpc of the Galactic disk is deficient 
with respect to that in the $R_{\rm GC}\gsim$~3~kpc region based on the
early CO observations and the MWISP data 
\citep[e.g.,][]{1975ApJ...202...30B,1977ASSL...70..165B,2001ApJ...547..792D,2021ApJ...910..131S}.
These results are consistent with the results from the large-scale
\HI\ and CO surveys \citep[see, e.g., reviews of][]{
1976ARA&A..14..275B,1990ARA&A..28..215D,1991ARA&A..29..195C,
1992Burton,2001ApJ...547..792D,2009ARA&A..47...27K,2015ARA&A..53..583H}.

Moreover, \HI\ study displays a good correlation between 
the \HI\ voids and the Fermi bubbles, indicating the physical 
connection between them \citep[][]{2016ApJ...826..215L}.
Benefiting from the large-scale \HI\ surveys, further studies show that the kinematic 
features of the atomic gas are very likely related to the Milky Way nuclear wind 
\citep[e.g.,][]{2009ApJS..181..398M,2013ApJ...770L...4M,2016ApJ...826..215L,
2018ApJ...855...33D,2020ApJ...888...51L,2020Natur.584..364D}.
Some authors also suggest that the large-scale multiwavelength features 
can be explained by the interaction between the Galactic nuclear wind 
(i.e., the GC superbubbles) and the Galactic gaseous disk 
\citep[e.g.,][]{2017PASJ...69L...8S,2021MNRAS.506.2170S}.

In this paper, we present the result of high dynamic range CO observations 
toward the large-scale view of the inner Galaxy. The kinematic information 
of the gas allows us to construct the needed molecular cloud (MC) samples 
from the CO emission, which is very helpful to trace the structures of 
the gaseous disk of the Milky Way.
At $\gsim$260~pc (and even $\gsim$600~pc) above and below the Galactic plane, 
the enhanced CO emission is discovered to be concentrated in a narrow range of 
$l\sim18^{\circ}-22^{\circ}$ or $R_{\rm GC}\sim$~2.6--3.1~kpc 
for the Sun's distance to the GC of $R_{0}$=8.15~kpc 
\citep[e.g.,][]{Reid19}, 
indicating the interface between the Milky Way nuclear wind and the gaseous disk.
Furthermore, some cometary CO structures exhibiting the head 
toward the Galactic plane and the tail away from the plane show 
that the cold molecular gas is entrained in multiphase flows.
The multiphase outflows are probably driven by the warm/hot gas
from the Milky Way nuclear wind.

In Section 2, we describe the CO, \HI, and IR data used in the paper.
In Section 3, we discuss the results and then give a simplified picture
to explain the multiwavelength observations.
Finally, Section 4 gives a brief summary based on our new findings.

\section{CO, \HI, and IR Data}
The CO data are from the Milky Way Imaging Scroll Painting survey
\citep[i.e., the MWISP project; see details in][]{2019ApJS..240....9S}.
Briefly, we employed the CO emission in the region of 
$l=12^{\circ}$--$26^{\circ}$ and $|b|\lsim5\fdg1$ to investigate the
molecular gas distribution near the tangent points, 
where the gas's distance is well determined. 
The spatial and spectral 
resolutions of the CO data are $\sim50''$ and $\sim$~0.2~$\km\ps$, 
respectively. After fitting the baseline and calibrating the main-beam efficiency, 
the reduced 3D data cubes (i.e., the position-position-velocity space, hereafter PPV)
with a grid spacing of 30$''$ have a typical rms 
noise level of $\sim$~0.5~K for 
\twCO\ ($J$=1--0) and $\sim$~0.3~K for 
\thCO/C$^{18}$O ($J$=1--0) at a channel width of 0.16~$\km\ps$. 
We mainly focus on the \twCO\ ($J$=1--0) emission in this paper.

In this work, the \HI\ data from the all-sky \HI\ survey 
\citep[][]{2016A&A...594A.116H} are used for large-scale comparisons
with the MWISP CO data.
The angular and velocity resolutions of the \HI\ data are 
16\farcm2 and 1.5~$\km\ps$, respectively. The typical RMS sensitivity
of the HI4PI data is $\sim$43~mK.
We also use the data from the survey of the Wide-field Infrared 
Survey Explorer \citep[WISE;][]{2010AJ....140.1868W} to investigate
the correlation between the molecular gas distribution and the dust emission.
The 12~$\mu$m and 22~$\mu$m WISE data used here have 
a spatial resolution of 6\farcs5 and 12\farcs0, respectively.

\section{Results and Discussions}
\subsection{Thinner Gaseous Disk within $R_{\rm GC}<$3~kpc}
The \HI\ data have shown that the atomic gas layer is noticeably thin in the 
region of $R_{\rm GC}\lsim$~3~kpc, probably indicating the interaction
between the Milky Way nuclear wind and the gaseous disk 
\citep[][]{1984ApJ...283...90L,2016ApJ...826..215L}.
And the early CO observations also show that 
the cold compressed gas is confined to a thinner layer within $R_{\rm GC}<$3~kpc
\citep[e.g.,][]{1975ApJ...202...30B,1976ARA&A..14..275B,1977ASSL...70..165B}.
Recently, a large number of small and high-velocity \HI\ clouds are found to be 
far above and below the disk toward the GC, suggesting that the large-scale atomic gas 
is physically associated with the Fermi bubbles and then the Milky Way nuclear wind
\citep[][]{2013ApJ...770L...4M,2018ApJ...855...33D,2020ApJ...888...51L}.

Furthermore, \cite{2020Natur.584..364D} have revealed that the cold and dense
molecular gas far from the plane can even survive in the hot and shocked nuclear wind.
These results show that energetic processes in the GC have profound effects on 
the distribution and evolution of the interstellar medium (ISM) on a large scale, 
which is consistent with the observational and theoretical studies on
other galaxies \citep[see the summary in][]{2020A&ARv..28....2V}.

As a widely used tracer of the MCs, CO data can provide 
the large-scale spatial and kinematic information of the H$_2$ gas. 
In this study, we explore the distribution and properties
of the molecular gas far from the Galactic plane based on the MWISP CO data.
The study is limited to the CO emission near the tangent points, which can avoid the
distance ambiguity and thus decrease the foreground confusion of the unrelated structures
\citep[see, e.g., the schematic view of Figure~1 in][]{2021ApJ...910..131S}.

We have confirmed that the inner molecular disk consists of two components:
the thin CO layer with a thickness of $\sim$85~pc and the thick layer
with $\sim$280~pc \citep[][]{2021ApJ...910..131S}.
The well-known thin CO disk harbors the majority of the molecular gas
in the Milky Way, while the thick CO disk is composed of many
small clouds in relatively high-$z$ regions. 
Some MCs in the thick CO layer are probably related to the feedback of 
the energetic star-forming activities near the Galactic plane
\citep[see, e.g., the case of microquasar SS~433 in][]{2018ApJ...863..103S}.
Interestingly, the two molecular gas layers are both thinner
within the $R_{\rm GC}\lsim$~3~kpc region \citep[i.e., an FWHM of $\sim$60~pc 
and $\sim$150~pc for the thin and thick CO layers, respectively;
see Figures 3-5 in][]{2021ApJ...910..131S}.

Figure~\ref{tangent} shows a large-scale distribution of the atomic and
molecular gas toward the tangent points. That is, we just integrated the gas
emission with velocities greater than the terminal velocities.
Here the terminal velocities can be determined from
the most positive velocity of the CO emission in the first quadrant
of the Milky Way \citep[see details in][]{2021ApJ...910..131S}.
Obviously, the disk traced by the atomic and molecular gas is indeed
thinner in regions of $l\lsim 22^{\circ}$ (or $R_{\rm GC}\lsim$~3~kpc).

Is the thinner gaseous disk within $R_{\rm GC}\lsim$~3~kpc related to 
the large-scale \HI\ voids or the Milky Way nuclear wind?
Is a substantial amount of cold neutral gas entrained by the Milky Way nuclear wind? 
If so, why is the neutral gas not destroyed by the high-velocity wind
in such harsh environments?
And what is the relation between the cool outflows and the hot wind from the 
nuclear region of the Milky Way?
The MWISP CO data with a wide spatial dynamic range, in combination with 
\HI\ and other multiwavelength observations, can give some hints on these topics.

\subsection{Enhanced high-$z$ CO Emission toward $l\sim$19\fdg1--22\fdg5}
MCs far from the Galactic plane may reveal some features of the large-scale 
structures related to the Milky Way nuclear wind.
To search for the high-$z$ MCs in the MWISP 3D datacube,
we use the density-based spatial clustering of applications with noise 
(DBSCAN\footnote{https://scikit-learn.org/stable/auto\_examples/cluster/plot\_dbscan.html}) 
clustering algorithm.
The algorithm is useful to identify CO clouds with noise in a big dataset,
which is crucial to reveal possible unusual features traced by weak emission.
Full details of the algorithm can be found in our previous studies
\citep[e.g.,][]{2020ApJ...898...80Y,2021ApJ...910..131S},
and a brief description of the method is shown below.

In order to further improve the signal-to-noise ratio, the MWISP raw data were resampled to 
0.5$\km\ps$, corresponding to a typical rms noise level of $\sim$~0.3~K per channel 
for the \twCO\ emission. For the PPV space, the minimum cutoff is 2$\times$rms and 
MinPts is set to 4 in the DBSCAN algorithm.
Then, post-selection criteria (i.e., $T_{\rm peak}\geq 4\times$rms, 
$\Delta l\times\Delta b\geq$~1~beam, $\Delta v\geq$~3~channels, and
the minimum number of neighborhood voxels $\geq$ 16) 
are used on the samples to remove the noise contamination. 
Here $T_{\rm peak}$ is the intensity per channel in the units of K, 
$\Delta l$ ($\Delta b$) is the spatial size in the units of arcmin, 
and $\Delta v$ is the spectral extension in the units of $\km\ps$.
Note that we use $T_{\rm peak}\geq 4\times$rms as the post-selection
criteria, which allows us to detect weaker CO emission in
the whole data compared to our previous studies
\citep[e.g., $T_{\rm peak}\geq 5\times$rms in][]{2021ApJ...910..131S}.

Furthermore, we identified the high-$z$ MCs toward the tangent points 
according to the selection criteria of $v_{\rm MC}\gsim v_{\rm tan}(l)-7$~km~s$^{-1}$
and $|z_{\rm MC}(l,b,v)| \geq 110$~pc 
\citep[i.e., $\geq3\times\sigma_z$ of the thin CO disk; see Table 3 in][]{2021ApJ...910..131S}. 
Here $v_{\rm MC}$ and $v_{\rm tan}(l)$ are the velocity of the MC and 
the corresponding tangent velocity at a certain longitude, respectively. 
And $\sigma_z$ is the standard deviation of the vertical distribution of 
the CO emission. By considering the cloud-cloud velocity dispersion of MCs 
\citep[][]{Malhotra94,2021ApJ...910..131S},
we adopted the value of $v_{\rm tan}(l)-7$~km~s$^{-1}$ to build
the high-$z$ MC samples near the tangent points. 

To reduce striping artifacts and other uncertainties in the whole data,
all selected MC samples are manually checked based on the cloud's spatial and 
velocity features in the PPV space. We find that the procedure is efficient
and valid. The method with the improved criteria can substantially increase 
the MC samples with weak CO emission, which is important to reveal 
the molecular gas distribution traced by small and faint clouds 
at the marginal signal-to-noise ratio of the MWISP survey.
And some interesting structures with weaker CO emission are indeed revealed
in our subsequent analysis (e.g., small MCs with $T_{\rm peak}\sim$1~K, see Section 3.3).

In total, 321 MCs near the tangent points were identified as the high-$z$ 
MC samples (i.e., $|z|\gsim$~110~pc) in the region of $l=12^{\circ}$--$26^{\circ}$ 
and $|b|\lsim5\fdg1$ (see purple circles in Figure~\ref{tangent}).
These discrete high-$z$ MCs usually have small sizes 
($\sim$0\farcm7--2\farcm2 or $\sim$1--5 pc in the radius), 
weak emission ($\sim$1--3~K in the peak temperature), and
high virial parameters ($\sim$5--30), which are consistent with
our previous studies \citep[e.g.,][]{2021ApJ...910..131S}.

Obviously, the high-$z$ MCs are well coincident with the distribution of 
the \HI\ emission near the tangent points, indicating the physical association 
between them (Figure~\ref{tangent}).
We check the spectral properties between the high-$z$ CO cloud and 
the \HI\ gas at the same location.
Both the CO cloud and the \HI\ gas have comparable LSR velocity near 
the tangent points, confirming the association of them.

Figure~\ref{lb} displays the 321 high-$z$ MCs in the $R_{\rm GC}$--$z$ space.
Besides the thinner gas layers discussed in Section 3.1, interestingly enough,
we also found that the extreme high-$z$ MCs (hereafter EHMCs) 
are located in two NARROW regions of [$l\sim$19\fdg1 to 20\fdg5, $b\sim$2\fdg0 to 5\fdg1] 
and [$l\sim$20\fdg5 to 22\fdg1, $b\sim-$2\fdg0 to $-5$\fdg1]
(see Figures~\ref{tangent} and \ref{lb}).
These EHMCs (i.e., $|z|\gsim$260~pc or $\sim3\times$FWHM 
of the thin CO layer) are unusual in such the narrow region 
far above and below the Galactic plane. 
The molecular gas mass of the EHMCs is also concentrated in the narrow range of
$R_{\rm GC}$=2.6--3.1~kpc (i.e., $\gsim9.2\times10^{3}\Msun$ in regions of 
$\sim 2\times \Delta R_{\rm GC} \times \Delta z$=$2\times 220$~pc$\times 410$~pc, 
see the histogram in Figure~\ref{lb} and Section 3.3).

These EHMCs, together with other high-$z$ MCs at $R_{\rm GC}\lsim$~2.3--2.6~kpc,
constitute molecular crater-wall structures lying along the edges of the
\HI\ voids (i.e., from $l\sim17^{\circ}$ and $|b|\gsim1^{\circ}$
to $l\sim22^{\circ}$ and $|b|\gsim5^{\circ}$, see Figure~\ref{tangent}).
We suggest that the molecular crater walls, in combination with the \HI\ voids 
above and below the Galactic plane, are related to the GC superbubbles
seen in radio, X-ray, and $\gamma$-ray emission 
\citep[e.g., Fermi bubbles;][]{2010ApJ...724.1044S}.

\subsection{EHMCs and Cometary Structures}
Table~1 lists the parameters of each EHMC:
(1) the ID of the EHMCs, arranged from the low Galactic longitude;
(2) and (3) the EHMC's Galactic coordinates ($l$ and $b$);
(4) the EHMC's LSR velocity ($v_{\rm LSR}$);
(5) the EHMC's one-dimensional velocity dispersion ($\sigma_v$);
(6) the EHMC's peak emission ($T_{\rm peak}$);
(7) the EHMC's effective radius (i.e., $d\times\sqrt{(\theta_{\rm MC}^2-\theta_{\rm beam}^2)/\pi}$, 
where $\theta_{\rm MC}$ and $\theta_{\rm beam}$, in units of arcmin, are the angular 
size of the CO emission and the beam size, respectively);
(8) the EHMC's distance obtained from the tangent points, i.e, $d=$8.15~kpc$\times$cos($l$)/cos($b$);
(9) the EHMC's $z$ height defined as $z=d\times$sin($b$);
(10) the EHMC's mass estimated from the CO-to-H$_2$ conversion factor method,
$X_{\rm CO}=2\E{20}$~cm$^{-2}$(K~km~s$^{-1})^{-1}$ \citep[e.g.,]
[]{2001ApJ...547..792D,2013ARA&A..51..207B};
and (11) the EHMC's virial parameter $\alpha=\frac{M_{\rm virial}}{M_{\rm Xfactor}}
=\frac{5\sigma_{v}^2R}{GM_{\rm Xfactor}}$,
where $R$ is the effective radius of the EHMCs and $G$ the gravitational constant.

Generally, the 47 EHMCs also have small sizes of several arcminutes 
(the mean value of 3\farcm1 and the median value of 2\farcm4),
weak emission (the mean value of 1.6~K and the median value of 1.8~K), 
and molecular masses spanning 40--3100~$\Msun$ 
(the mean value of 340~$\Msun$ and the median value of 110~$\Msun$).
Based on the new criteria (Section 3.2), about half of the EHMCs 
are small clouds with a mass of $\lsim 100\ \Msun$.
As the location anchors, the EHMCs are essential to trace the molecular 
crater-wall structures located along the edges of the \HI\ voids 
at $\sim18^{\circ}$--$22^{\circ}$ (Figure~\ref{tangent}).

Moreover, some large EHMCs display spatially resolved structures.
Two zoomed-in views of the EHMCs 
are shown in Figure~\ref{twosamples} for EHMC G019.957$+$02.863 and 
EHMC G021.548$-$03.414 (also see regions of the red boxes in Figure~\ref{tangent}).
Based on the new MWISP CO data
at moderately high angular resolution and fairly high sensitivity,
we find that the two EHMCs traced by CO emission (black contours) 
display the elongated head-to-tail structures pointing away from the 
Galactic plane (see blue arrows in Figure~\ref{twosamples}).
We also find that the elongation of the molecular gas coincides exactly with 
the atomic gas ridges revealed by \HI\ emission (see purple contours in Figure~\ref{twosamples}) 
and the dust filaments traced by IR emission. 

Remarkably, the cometary structure of the EHMC G021.548$-$03.414 
can be clearly seen from the CO emission (i.e., see CO black contours
of the $\sim$~20~pc long structure in the bottom panel of Figure~\ref{twosamples}).
The CO emission of the head of the EHMC is brightest toward the Galactic plane,
while the cometary tail is faint away from the plane.
At the head of the EHMC, the \twCO\ peak temperature is $\sim$~4.4~K, 
which is about four times of the detected \thCO\ emission at the brightest part of 
the \twCO\ emission (i.e., $T_{\rm peak13}\sim4\times$rms13=1.1~K, see Table~1). 
We also show the peak temperature of the detected \thCO\ emission for the EHMCs in Table~1.
No C$^{18}$O emission is detected for any identified EHMCs.

Contrary to the bright CO emission at the head of the cometary cloud, 
the IR dust emission is bright in the tail regions where the atomic gas is also 
enhanced based on the \HI\ data with a grid of $5'$ (Figure~\ref{twosamples}).
The feature probably indicates that the entrained molecular gas (and dust) 
in the tail is heated by the surrounding warm/hot gas, leading to the multiphase flows 
from the cold head to the warm tail in the crater-wall region.
Based on the WISE data, the dust temperature at the tail of 
the EHMC G021.548$-$03.414 is probably $\sim$100--500~K, which is larger 
than the molecular gas temperature at the head of the cloud 
(e.g., the estimated molecular gas temperature of $\lsim$~10~K based on 
the optically thick \twCO\ emission and the beam filling factor of 1). 
We must stress that the derived dust temperature is dependent on
the accurate IR flux from the thermal emission at several bands. 
More observations are necessary to draw a solid conclusion.
On the other hand, there are at least three MC structures 
(i.e., EHMC ID 13, 14, and 15 in Table~1)
concentrated in the region toward EHMC G019.957$+$02.863.

Figure~\ref{pv} shows the position--velocity (PV) diagrams of the two
EHMCs from the head to the tail (see blue arrows in Figure~\ref{twosamples}).
For EHMC G021.548$-$03.414, the velocity gradient of $\sim -0.23\km\ps$pc$^{-1}$
is found to be from the denser head to the faint tail.
Considering a small inclination angle of $i\sim10^{\circ}-20^{\circ}$ 
for the outflows on the sky, the true velocity gradient could be larger 
(e.g., $\sim -1\km\ps$pc$^{-1}$ for the moving flows roughly perpendicular 
to the line of sight).
The entrained molecular gas of the EHMC is moving toward us 
based on the observed blueshifted CO emission.

These new findings indicate that the head of the high-$z$ MCs mainly 
contains cold and dense molecular gas, while the extended filamentary tail 
incorporates a substantial amount of atomic gas, molecular gas, 
and dust stripping from the MC's head.
That is, the molecular gas is changing phase to become the atomic gas 
(and very likely the warm ionized gas) as it moves to the high-$z$ regions.
The material in the tail is comoving with the surrounding 
warm/hot ionized gas from the low-$z$ regions to the high-$z$ regions, 
providing a supply of fresh gas to the Milky Way halo and eventually 
falling back onto the plane (e.g., the velocity of cool outflows at 
$\sim140-330\km\ps$ is less than the escape velocity of the Milky Way;
see Section 3.4.2).

The association of the bright dust emission with the cool gas in the tail
shows that the dust can survive long in the crater walls with
the local high density (e.g., at least several Myr; see Section 3.4.3).
The survival of dust is very important 
for efficient formation of molecules in the interface between 
the cool outflows and the warm/hot wind.
The observation features are also consistent with the recent simulation results
that the molecular and dusty clouds are likely to survive long enough
in warm/hot winds \citep[e.g.,][]{2018MNRAS.480L.111G,2020MNRAS.492.1970G,
2019MNRAS.486.4526B,2022MNRAS.510..551F}.

Some other elongated high-$z$ CO clouds lying in the edges of the
\HI\ voids are also shown in Figure~\ref{lb4s}. 
We note that the MCs in panels (c) and (d) of Figure~\ref{lb4s} form
a $\sim$~140~pc long structure from $b\sim-$2\fdg6 to $b\sim-$3\fdg5,
supporting the existence of the large-scale molecular crater walls
along the edges of the \HI\ voids (and the boundary of the GC superbubbles;
see Figure~\ref{tri}).

\subsection{Impact of Galactic Winds on the Gaseous Disk}
\subsubsection{Origin of the EHMCs}
There are probably two scenarios that can explain the observed EHMCs far from the Galactic
plane, i.e., (1) the local star-formation-feedback scenario
and (2) the Milky Way nuclear wind scenario. 
We will discuss the two scenarios described below.

For the local star formation feedback scenario,
the feedback from star-forming activities near the Galactic plane has profound 
effects on the surrounding ISM environment, i.e., changing the physical properties and 
the distributions of the gas. 
The energetic processes from massive stellar winds and/or supernova explosions can 
produce large-scale structures such as supershells, superbubbles, and large-scale chimneys. 
Especially, the region between $l\sim20^{\circ}$ and $30^{\circ}$ is close to the near tip
of the Galactic bar and it hosts intense star-forming regions in the Galactic plane.
The region also shows a rich population of \HI\ 
extraplanar clouds that are not seen in other regions of the Galaxy
\citep[the \HI\ scale height in the region of $l\sim20^{\circ}$
is twice the scale height of the corresponding region in $l\sim -20^{\circ}$, see
][]{2008ApJ...688..290F,Ford10}

Therefore, the high-$z$ MCs ($|z|\gsim$~100~pc) in the thick CO disk may be the debris of 
the disk gas that was blown away by local star-formation feedback near the Galactic plane 
\citep[see, e.g., discussions in][]{2021ApJ...910..131S}. For example,
the identified high-$z$ MCs at ($l\sim$40\fdg3, $b\sim -$4\fdg3) are probably associated with 
the nearby SS~433/W50 system \citep[e.g., $z_{\rm MC}\sim-$400~pc at a distance of 5~kpc; see]
[]{2018ApJ...863..103S}.
Is it possible that the EHMCs at $l\sim20^{\circ}$ are from the energetic processes 
of stellar feedback near the Galactic plane, for example, the Ophiuchus superbubble 
\citep[see details in][]{2007ApJ...656..928P}?

Based on results from the MWISP data, we find that the high-$z$ MCs with very faint 
CO emission are highly turbulent, indicating that 
the cold gas is mixing with the warmer gas and the MCs are either dispersing or being
assembled by external dynamical processes. These small-sized (1--4~pc) high-$z$ MCs
are probably short-lived, e.g., less than their typical internal crossing time of several Myr.
The clouds thus cannot move too far from the disk if they are directly from the 
Galactic midplane, conforming to the observed MCs' distribution for the thick CO
disk \citep[e.g., $\sigma_z\sim$110--120~pc; see][]{2021ApJ...910..131S}.
Here we ignore the case that the high-$z$ molecular gas may condense in situ.
Indeed, we find no detailed correlation between the Ophiuchus superbubble in \HI\
emission and the high-$z$ MCs near the tangent point in a region of
$l\sim20^{\circ}$--$40^{\circ}$ and $z$=100--600~pc.
On the other hand, the Ophiuchus superbubble appears to be one-sided and does not
extend below the Galactic plane \citep[e.g.,][]{2007ApJ...656..928P}, 
although the HI4PI data show a lot of anomalous structures below 
the Galactic plane \citep[see also][]{Ford10,2021ApJ...910..131S}. We 
argue that the enhanced EHMCs at $l\sim18^{\circ}$--$22^{\circ}$
(or $R_{\rm GC}\sim$2.6--3.1~kpc) are not associated with the old Ophiuchus superbubble
\citep[e.g., an age of about 30~Myr; see][]{2007ApJ...656..928P}. 

In principle, some EHMCs are probably related to the young star-formation feedback 
(e.g., several Myr) near the Galactic plane. However, the extended direction of the 
cometary EHMCs (see Figure~\ref{twosamples}) and the coherent EHMC samples in Figure~\ref{tri} 
show that the large and cometary EHMCs are likely related to dynamical processes toward 
regions of $l\lsim18^{\circ}$--$22^{\circ}$ (or toward the GC direction), 
in which the star-forming activities near the Galactic plane 
are not very intense except for the Central Molecular Zone (CMZ) region.

According to the above discussions, we propose that most of the EHMCs 
in $R_{\rm GC}\sim$2.6--3.1~kpc are associated with the Milky Way nuclear wind. 
We also summarize the observation results as below:\\
(1) The dominant EHMCs are just located near the edges of the large-scale \HI\ 
voids toward $l\sim18^{\circ}-22^{\circ}$ or $R_{\rm GC}\sim$~2.6--3.1~kpc 
(see Figures~\ref{tangent}, \ref{lb}, and \ref{tri}), indicating the association 
between them.\\ 
(2) The large EHMCs along the edges of the \HI\ voids display 
unusual head-to-tail structures pointing away from the Galactic plane 
(or the direction away from the GC; see Figures~\ref{twosamples} and \ref{lb4s}).\\
(3) The observed velocity gradient of EHMC G021.548$-$3.414, together with its 
large velocity width (Figure~\ref{pv}), supports the idea that the MC is unstable 
and is partially destroyed by ambient dynamical processes. 
That is, the molecular gas is accelerated from the head to the tail 
owing to the blueshifted CO profile for the tail structure (see the bottom
panel of Figure~\ref{pv}).
The feature supports the entrainment scenario that the material is moving from the 
Galactic plane to the high-$z$ regions.\\
(4) The IR emission of EHMC G021.548$-$3.414 (Figure~\ref{twosamples}) is bright at 
its long tail but is faint at the dense head, favoring the scenario of the molecular gas 
being ablated and heated from the cloud edges by a warm/hot wind.

These observational features can be naturally explained by the entrainment scenario
that the cold gas is moving with the multiphase medium at the interface between
the warm/hot nuclear wind of the Milky Way and the gaseous disk.
The process also leads to the concentrated molecular gas (and mass) in narrow
regions of $l\sim18^{\circ}-22^{\circ}$ (or $R_{\rm GC}\sim$2.6--3.1~kpc) and 
$|z|\gsim 260$~pc (see the cometary EHMCs along the edges of the \HI\ 
voids in Figures~\ref{twosamples} and \ref{tri}, $|z|\sim 400-450$~pc$\gsim 3 \sigma_z$
of the thick CO disk).

The \HI\ voids, which are physically associated with the Fermi bubbles,
are thus surrounded by enhanced CO emission at low latitudes of
$|b|\gsim2^{\circ}$--$5^{\circ}$ and $R_{\rm GC}\lsim$2.6--3.1~kpc.
This scenario also agrees well with the recent observation that the Milky Way nuclear wind
can remove gas from the disk to the halo \citep[e.g.,][]{2021ApJ...923L..11C}
and can accelerate MC fragments into the high-$z$ regions
\citep[e.g.,][]{2020Natur.584..364D}.

Finally, note that not all EHMCs in Table~1 are related to the Milky Way nuclear winds. 
Only samples of ID 05--31 in Table~1 are used to calculate the mass of the 
crater walls (see Figure~\ref{lb}).
Among the EHMCs in the crater walls, over 60\% of samples are resolved and 
nearly 30\% of EHMCs (ID 13, 14, 15, 19, 23, 24, and 27) display cometary structures.
These features are unusual in the narrow region of $R_{\rm GC}\sim$2.6--3.1~kpc.
And the dominant molecular mass in the CO crater walls is from the large-size 
and cometary EHMCs that are located along the edges of the \HI\ 
voids (e.g., $R_{\rm GC}\sim$2.6--3.1~kpc; see EHMCs in Figure~\ref{tri} 
and the caption). 
Therefore, the possible contamination from the local star-formation feedback 
near the Galactic plane has little effect on the subsequent estimation assuming 
that the cometary and large EHMCs along the crater walls are (very likely) 
of Milky Way nuclear wind origin.

Of course, observing a similar region at the other side of the crater 
(i.e., $l\lsim -20^{\circ}$), where star-formation activity is not very intense, 
would be a neat confirmation.
However, the MWISP survey cannot cover that longitude range.

\subsubsection{Cool Outflows in the Milky Way}
Multiphase outflows are a common feature in galaxies \citep[e.g.,][]{2020A&ARv..28....2V}. 
For the Milky Way, multiwavelength observations have revealed many interesting
large-scale structures related to the Galactic nuclear wind, 
i.e, the GC Lobes in radio \citep[e.g.,][]{1984Natur.310..568S,2013Natur.493...66C}, 
IR \citep[e.g.,][]{2003ApJ...582..246B}, X-ray \citep[e.g.,][]{2020Natur.588..227P}, 
$\gamma$-ray \citep[i.e., the Fermi bubbles in][]{2010ApJ...724.1044S} emission 
and UV absorption \citep[e.g.,][]{2008ApJ...679..460Z,2015ApJ...799L...7F,
2017ApJS..232...25S,2018ApJ...860...98K,2020ApJ...898..128A}, 
showing the bipolar outflows on scales of a few degrees to tens of degrees. 

New multiwavelength analyses also found chimney-like structures that 
are physically related to the intermittent activity near the GC 
\citep[][]{2019Natur.567..347P,2021A&A...646A..66P}.
Recently, the \HI\ studies have revealed spatial
\citep[e.g., \HI\ holes in][]{2016ApJ...826..215L} and kinematic atomic gas features
\citep[e.g., anomalous high-velocity clouds extending up to the high-$z$ regions;]
[]{2013ApJ...770L...4M,2018ApJ...855...33D,2020ApJ...888...51L} 
associated with the Galactic wind.
These studies support the existence of large-scale multiphase outflows in our Galaxy
(i.e., the neutral gas at $T\lsim 10^{2}$~K, warm ionized gas $T\sim 10^{3}-10^{4}$~K, 
and high-temperature gas at $T\gsim 10^{6}$~K).

Based on the discussions in Section 3.4.1, we suggest that large 
amounts of molecular gas, which is concentrated in the edges of the \HI\ 
voids associated with the Fermi bubbles, is entrained in large-scale 
multi-phase outflows from the Galactic gaseous disk to the high-$z$ regions.

The detected EHMCs could be an important mass reservoir of the cool outflows
in the Milky Way. The molecular mass in the crater-wall structures 
of $l\sim$19\fdg1--22\fdg5 and $|z|\gsim$~260~pc 
(i.e., Area$\sim 2\times \Delta R_{\rm GC} \times \Delta z$=
$2\times 220$~pc width $\times410$~pc height; 
see Figure~\ref{lb} and ID 05--31 in Table~1) is estimated to be 
$\gsim9.2\times10^{3}\Msun$ by adopting the CO-to-H$_2$ conversion factor of
$X_{\rm CO}=2\E{20}$~cm$^{-2}$(K~km~s$^{-1})^{-1}$ \citep[e.g.,]
[]{2001ApJ...547..792D,2013ARA&A..51..207B}.
In the meanwhile, many small clouds with weak CO emission (e.g., $T_{\rm peak}\lsim$~1~K) 
may exist in the walls of the crater-like structures, but we cannot pick them 
out owing to the beam dilution and the limited sensitivity of the survey data.
For a conservative estimate, the mean volume density of the molecular gas 
in the crater walls is $\gsim3\times10^{-4}\ \Msun$~pc$^{-3}$ 
(or $\gsim0.01$~H~cm$^{-3}$) 
assuming a thickness of $\sim$200~pc along the line of sight (LOS).

Due to confusion with the unrelated \HI\ emission near the Galactic plane,
the total mass of the atomic gas cannot be precisely measured in the same
region of the EHMC concentrations.
Based on the \HI\ emission near the tangent points, however, 
the upper limit of the total mass of the atomic gas 
can be roughly estimated to be $\lsim1.3\times10^{5}\ \Msun$
from I$_{{\rm H\textsc{i}}}(v_{\rm LSR}>v_{\rm tan}$),
leading to the atomic gas density of $\lsim4\times10^{-3}\ \Msun$~pc$^{-3}$
(or $\lsim0.1$~H~cm$^{-3}$). Note that the estimated \HI\ mass contains
the attribution from some unrelated gas structures with $v_{\rm LSR} \textless  v_{\rm tan}$
because of the broad line width of the \HI\ emission.
Therefore, the derived atomic gas mass (and the density) toward 
the tangent points is probably the upper limit based on the 
estimated value of $I_{{\rm H\textsc{i}}}(v_{\rm LSR} \textgreater v_{\rm tan}$).

The mean density of the cold gas estimated above is at least one order of 
magnitude larger than the hot gas density in the bubbles 
\citep[e.g., the often-used value of $\sim10^{-3}$~H~cm$^{-3}$ in]
[]{2003ApJ...582..246B,2015ApJ...808..107C,2015MNRAS.453.3827S,
2016ApJ...829....9M}, suggesting that the hot gas in the nuclear wind 
is surrounded by dense and cold shells at the boundaries of 
the Fermi bubbles near the Galactic gaseous disk. 
Therefore, the cold gas at $R_{\rm GC}\sim$3~kpc likely confines 
the hot wind near the gaseous disk at least on the height of 
$|z|\sim$600~pc (see Figures~\ref{tangent} and \ref{lb}).
That is what we observed based on the MWISP CO survey 
and the combination of the \HI\ data.

Considering the bipolar-outflow structures for the Milky Way nuclear wind, 
we can estimate that the total mass of the molecular crater walls
should be $\gsim1\times10^{6}\ \Msun$ for regions of
$\sim 2\times 2\pi R_{\rm GC} \times \Delta R_{\rm GC} \times \Delta z$
=$3.4 \times 10^{9}$~pc$^{3}$
(i.e., the two bowl-like structures above and below the Galactic plane 
at $R_{\rm GC}\sim$3~kpc and $|z|\sim$260--670~pc).
It is interesting to note that the total molecular mass in the crater walls
is well comparable to that of the \HI\ gas in the Fermi bubbles
\citep[i.e., $\sim10^{6}\ \Msun$ in][]{2018ApJ...855...33D,2020ApJ...888...51L}.
Additionally, some molecular gas could still survive in the inner of the 
Fermi bubbles \citep[e.g.,][]{2020Natur.584..364D},
which will increase the total molecular mass of the cool outflows 
associated with the Milky Way nuclear wind.

Our results show that a large amount of neutral gas 
(the total atomic and molecular gas mass of $\sim 10^{7}\ \Msun$ at $|z|\gsim$~260~pc) 
is located in the crater walls,
which surround the base of GC superbubbles at low latitudes
above and below the Galactic plane.
The high-$z$ molecular gas, together with the related atomic gas and dust,
constitutes the cool outflows associated with the Milky Way nuclear wind.

Figure~\ref{tri} shows large-scale velocity distributions of the gas 
along the crater-wall structures (black dashed lines in the left panel).
We find that the crater-wall structures display the systematic
velocity gradient of $\sim -0.03\km\ps$pc$^{-1}$ along the LOS.
That is, the observed velocity along the LOS decreases with increasing 
the height of $|z|$, for regions both above (e.g., EHMC IDs~5--13 in Table~1) 
and below (EHMC IDs~19--28 in Table~1) the Galactic plane.
The lag is probably the result of the interaction between the entrained gas
from the disk and the slowly rotating gas in the halo
\citep[e.g.,][]{2008MNRAS.386..935F,2009MNRAS.399.1089M,2015MNRAS.451.4223M,2016ApJ...826..215L}.

Additionally, two substructures, which are identified from the
coherent EHMCs with similar spatial and velocity features in the crater walls, 
display the positive velocity gradient with increasing the height of $|z|$
(see the black solid lines in the right panels of Figure~\ref{tri}).
Here the coherent EHMCs mean that (1) the clouds have similar LSR 
velocities in a small region and (2) they have similar elongations along the 
crater walls (or the edges of the \HI\ voids).
Both the substructures traced by CO emission are located at $\sim$400--450~pc 
far from the Galactic plane (see, e.g., black contours in Figure~\ref{twosamples}).

The velocity gradients of the two substructures are
$\sim 0.15\km\ps$pc$^{-1}$ for the $\sim$~20~pc long structure above the plane 
(see the black solid line in the top right panel of Figure~\ref{tri} 
for EHMCs IDs 13, 14, 15, and 17 from Table~1) and 
$\sim 0.16\km\ps$pc$^{-1}$ for the $\sim$~40~pc long one below the plane 
(see the black solid line in the bottom right panel of Figure~\ref{tri}
for EHMCs IDs 23, 24, 26, and 27 from Table~1), respectively.
The true velocity gradient is likely 3--6 times larger than the observed gradient
along the LOS by considering the projection correction of
the small inclination angle 
(e.g., $\nabla v_{\rm true}=\nabla v_{\rm obs}$/sin($i$) 
for $i\sim10^{\circ}-20^{\circ}$; see Section 3.3).

We argue that the velocity gradient of the large-scale coherent EHMCs probably 
results from cool outflows associated with the Milky Way nuclear wind.
The velocity of cool outflows is roughly $\sim 140-330\km\ps$
(i.e., $v_{\rm w}\sim path\times \nabla v_{\rm true}$)
assuming that the CO gas comes from the locations of 110--130~pc far from
the Galactic plane at a constant acceleration 
(e.g., $path\sim$~280--330~pc and $\nabla v_{\rm true}\sim0.5-1\km\ps$pc$^{-1}$).
Here we tentatively assume that the entrained high-$z$ gas is launched from the 
boundary of the thin CO disk to their current places (see, e.g., the cometary CO 
structures at $b\sim 1^{\circ}$ or $|z|\sim$120~pc in the top panels of Figure~\ref{lb4s}).
The estimated velocity of the cool outflows is roughly
comparable to the value of $\sim$200--300$\km\ps$ from the \HI\ kinematic models
\citep[e.g.,][]{2013ApJ...770L...4M,2018ApJ...855...33D,2020ApJ...888...51L}.

The cold gas in the multiphase outflows is mainly entrained 
along the walls of the hot gas cavity blown by the Milky Way nuclear wind.
The scenario is also similar to the famous examples
of nearby starburst galaxy M82 \citep[e.g.,][]{2015ApJ...814...83L,2019PASJ...71...87Y,
2021ApJ...915L...3K} and NGC~253 \citep[e.g.,][]{2013Natur.499..450B,2015ApJ...801...63M,
2017ApJ...835..265W,2019ApJ...881...43K}.
Further observations and simulations are helpful in understanding
the whole picture of the Galactic multi-phase nuclear outflows/winds
\citep[e.g.,][]{2020ApJ...894....1F,2020ApJ...894..117Z,2021MNRAS.506.5658B,
2021ApJ...922..254C,2021MNRAS.508.4667P,2021arXiv210903834M,2022ApJ...924...82F,
2022AJ....163..134T,2022NatAs.tmp...52Y}.

\subsubsection{Survival of Molecular Gas in the High-$z$ Regions}
In principle, the isolated and cool clouds 
(e.g., $n\sim$10--1000~cm$^{-3}$ and $T\lsim 10^{2}$~K) will be eventually 
destroyed in the harsh environment (e.g., the high-velocity shock and/or 
the surrounding warm/hot winds, $n\sim 10^{-3}$--1~cm$^{-3}$ and $T\gsim10^{4}$~K).
The cloud-crushing time \citep[e.g.,][]{1994ApJ...420..213K} can be defined as 
$t_{\rm cc}=\chi^{\frac{1}{2}}\frac{2r_{\rm cloud}}{v_{\rm w}}$, 
where $\chi=\frac{\rho_{\rm cloud}}{\rho_{\rm wind}}$ is the density contrast 
between the cloud and the wind,
$r_{\rm cloud}$ is the cloud radius, and $v_{\rm w}$ is the wind velocity. 

By adopting $\chi=1000$ and $v_{\rm w} =200\km\ps=v_{\rm w200}$ (see Section 3.4.1),
we find that the crushing time of the high-$z$ MCs is $t_{\rm cc}\sim0.9v_{\rm w200}^{-1}$~Myr
for the typical EHMC radius of $r_{\rm cloud}$=3~pc.
The crushing time of the EHMCs is slightly small compared to the dynamical time of the
gas flows (i.e., $t_{\rm dyn}=\frac{l}{v_{\rm w}}\gsim1.3v_{\rm w200}^{-1}$~Myr, 
where the moving distance of gas flows is $l\gsim$260~pc for the EHMCs).
Note that changing $\chi=1000$ to $\chi=100$ will decrease the crushing time
by a factor of $\sim$3.
The smaller clouds (e.g., $r_{\rm cloud}\ll$~1~pc) thus could not 
survive long in the high-velocity wind.

For a more realistic case, however, EHMCs with parsec scales likely survive for a long 
period of time (e.g., a few Myr for several times of $t_{\rm cc}$) in the gas-rich 
environment by considering the radiative cooling, condensation of gas from warm clouds, 
and some other effects
\citep[see, e.g.,][]{2015MNRAS.449....2M,2017MNRAS.470..114A,2018MNRAS.480L.111G,
2020MNRAS.492.1970G,2022MNRAS.511..859G,2019MNRAS.486.4526B,2020ApJ...895...43S,
2020MNRAS.499.4261S,2021MNRAS.505.1083G,2021MNRAS.501.1143K,2022MNRAS.510..551F}.

As an interesting example, adopting the velocity gradient of $\sim 1\km\ps$pc$^{-1}$ 
for EHMC G021.548$-$03.414 with the projection correction (Section 3.3),
its dynamic time is estimated to be 
$t_{\rm dyn}=t_{\rm acc}=2\times\frac{\Delta length}{\Delta v}\sim2.0$~Myr, 
which is comparable to its crushing time of $t_{\rm cc}\sim2.2v_{\rm w200}^{-1}$~Myr
for its effective radius of 7.2~pc (see Table~1).
The long tail of the cloud, together with the revealed velocity gradient, 
indicates that the molecular gas is ablated by the surrounding high-velocity wind. 
The molecular gas of the cloud is entrained in the multiphase flows,
in which the molecular gas will be transformed to the neutral atomic 
and/or warm ionized gas moving toward the high-$z$ regions.
We suggest that the EHMC is crushed and will be destroyed in future several Myr.
Thus, the lifetime of the EHMC at $|z|\sim$450~pc far from the Galactic plane 
is probably $\sim$5--10~Myr, which is much longer than its crushing time.

Our CO observations reveal large reservoirs of cool gas surrounding the boundaries of
the large-scale nuclear wind near the gaseous disk.
That is, the abundant atomic gas is concentrated in the crater-wall structures that 
the dense EHMCs are embedded in (Sections 3.2 and 3.3).
The EHMCs thus are not isolated objects situated in an empty space. 
The gas-rich environment with the local high density increases the survival ability 
of MCs in the crater walls.

For example, mass loading from the gaseous disk to the crater walls 
can flatten the density and temperature profiles on the boundary of 
the wind bubbles, creating multiphase flows in such regions.
The multiphase gas-rich environment may cool fast to replenish the
cold gas reservoir and then extend the lifetime of the high-$z$ MCs.
Additionally, the newly cooled gas from the surrounding high-velocity flows
can carry momentum of the hot gas, leading to the
observed entrainment scenario in the porous and mixed multiphase medium.

Nearly no EHMC is observed in the regions of $R_{\rm GC}\lsim$2.6~kpc (Figure~\ref{lb}).
There are two plausible reasons for the feature.
One is that high-$z$ clouds are indeed destroyed by the hot winds,
in which the wind velocity in the nuclear wind bubble is higher 
than that near the boundary of the bubble.
The molecular gas of the clouds may be rapidly transformed to 
the warm/hot ionized gas (e.g., $t_{\rm cc}\lsim\ 0.2$~Myr for 
the high-velocity wind with $v_{\rm w}=1000\km\ps$ and $T\gsim10^{6}$~K). 
The ionized gas moves fast toward the high-$z$ regions, 
in which little cool gas can survive in the hot flows.
On the other hand, our EHMC samples are from the CO emission toward the tangent points.
The EHMC samples only occupy a small volume of the bubble in a certain LOS. 
It thus decreases the detection rate of possibly survived high-$z$ MCs
with an origin size of tens of parsecs in the hot wind bubble.

Finally, we emphasize that the cloud, and even the ISM, is
inherently complex in its structure rather than the isolated and homogenous
distribution in temperature, velocity, and density.
The mean volume density of the EHMCs is $\sim$20~H$_2$~cm$^{-3}$,
much below the CO critical density of 3000/$\tau_{\rm 12CO}$~H$_2$~cm$^{-3}$
\citep[see, e.g.,][]{2013seg..book..491S,2015PASP..127..299S}.
This probably shows that the EHMCs consist of clumpy and multiphase medium,
in which the highly structured molecular gas with a low volume filling factor
is mixed with more diffuse gas \citep[e.g.,][]{1991ApJ...378..186F,1992A&A...257..715F,
1996ApJ...472..191F,2006ARA&A..44..367S,2016A&A...591A.104H}.

The original material in the MCs is heated into the multiphase gas (i.e., ionized/atomic/molecular)
and entrained as it mixed with the warm/hot wind. Once the cloud material is entrained, 
it may quickly cool back down to the molecular phase in the enhanced gas+dust environment.
In such a scenario, many effects (e.g., conduction and cooling,
magnetic field and cosmic rays, and various instabilities) should be carefully
taken into account for the interaction between the fractal MCs and the warm/hot wind.
Detailed analysis of these effects is beyond the scope
of this paper. 
More multi-wavelength observations and simulations will be very helpful
to clarify these issues.

\subsubsection{Milky Way Nuclear Wind and the Gaseous Disk}
In the crater-wall regions (i.e., $|z|\sim260$--670~pc and $R_{\rm GC}\sim$3~kpc),
the total mass of the molecular gas is $\gsim1\times10^{6}\ \Msun$.
For the highest EHMC at $|z|\sim$620~pc, its dynamical time is estimated to be
$t_{\rm dyn}=\frac{620\ {\rm pc}}{v_{\rm w200}}\lsim3.0v_{\rm w200}^{-1}$~Myr.
The mass-loading rate from the inner gaseous disk of $R_{\rm GC}\lsim$3~kpc 
to the high-$z$ regions is thus $\gsim0.3v_{\rm w200}\ \Msun$~yr$^{-1}$.

If we take into account the disturbed MCs in the region close
to the Galactic plane (e.g., 110~pc$\lsim|z|\lsim$260~pc), 
the true mass-loading rate may be increased by a factor of $\sim10$
by assuming a Gaussian distribution 
\citep[e.g., the total mass is $\sim10$ times of that in $|z|\gsim$260~pc regions 
for the thick CO disk of $\sigma_z\sim$110--120~pc; see][]{2021ApJ...910..131S}.
By considering the neutral atomic gas coexisting with the molecular gas, 
the mass-loading rate of the cool outflows may be slightly larger than the 
estimated value of $\sim3v_{\rm w200}\ \Msun$~yr$^{-1}$. 
Although large uncertainty remains and more observations are needed,
the rough estimate shows that the outflow rate at the order of
$\sim2-4\ \Msun$~yr$^{-1}$ is possible according to the large-scale enhanced 
CO emission at $R_{\rm GC}\sim$3~kpc. 

The energy source of the Milky Way nuclear wind is still being debated,
i.e., intermittent activities from Sgr A* for AGN-like model vs. integrated 
effects of the stellar feedback from the CMZ for the starburst model.
The total kinetic energy of the cool-gas outflows can be estimated 
as $E_{\rm K}=0.5M_{\rm gas}v_{\rm w}^2\gsim4\times10^{54}v_{\rm w200}^{2}$~erg
for the total molecular gas mass of $\gsim1\times10^{7}\ \Msun$ at $|z|\gsim$110~pc
(e.g., about $\sim10$ times of that in $|z|\gsim$260~pc regions).
The required kinetic power to the pushed molecular gas is then
$\sim4\times10^{40}v_{\rm w200}^{3}$~erg~s$^{-1}$ for the assumed
dynamical time of $\sim3.0v_{\rm w200}^{-1}$~Myr.
Whatever the exact origin of the Milky Way nuclear wind, our estimates show that 
the cool-gas outflows can be easily powered by the energetic processes 
near the GC (e.g., the total energy of $10^{56}-10^{57}$~erg or 
the total power of $10^{42}-10^{44}$~erg~s$^{-1}$ for the Fermi bubbles).

On the other hand, low-density gas can be easily accelerated in the hot wind environment.
And the hot ionized gas is probably the dominant reservoir of energy of the Fermi bubbles
\citep[e.g., the inferred temperature of $\gsim2\times10^{6}$~K,
the low density of $\sim10^{-3}$~cm$^{-3}$, and the typical velocity of 
$\sim500-1000\km\ps$ for the hot gas;][]{2016ApJ...829....9M,2017ApJ...834..191B}.
The warm and hot ionized gas of the nuclear wind plays an important role 
in driving the cool-gas outflows near the gaseous disk. 
Accordingly, the ablated gas from the high-structured MCs 
in the gaseous disk joins the moving flow, modifying the velocity, 
temperature, and density distribution of the multi-phase medium.

The thinner disk of the atomic and molecular gas within the region of 
$R_{\rm GC}$\lsim3~kpc can be explained by the effect of 
the large-scale Milky Way nuclear wind. 
Assuming the hot wind velocity of $\sim500-1000\km\ps$ 
and the mass-lose rate of $\sim10-20\ \Msun$~yr$^{-1}$ 
near the Galactic inner gaseous disk, 
the disk will lose $\sim6\times10^{7}\ \Msun$ in a period of $\sim$3--6~Myr
(e.g., from the CMZ to the $R_{\rm GC}\sim$3~kpc regions), 
which is about 30\% of the molecular gas within the $R_{\rm GC}$\lsim3~kpc region
\citep[e.g., the total molecular mass of $\sim2\times10^{8}\ \Msun$, see][]
{2015ARA&A..53..583H,2016PASJ...68....5N}.
Therefore, a bulk of the molecular gas can be removed from the gaseous disk
within $R_{\rm GC}\lsim$3~kpc, especially for the region of $|z|\gsim$~50--100~pc.

The total mass of the removed molecular gas in the inner disk
is roughly comparable to the value of $\sim7\times10^{7}\ \Msun$ 
for the whole 3-kpc arm \citep[e.g., refer to H$_2$ masses per unit length of 
$\sim4\times10^{6}\ \Msun$~kpc$^{-1}$ in][]{2008Dame}.
We propose that the 3-kpc arm is probably related to the Milky Way nuclear wind.
It is true that only a small fraction of molecular gas ($\gsim1\times10^{6}\ \Msun$) 
is accelerated to the $|z|\gsim$260~pc region, while the dominant molecular gas seems to be accumulated 
at the disk \citep[e.g., the 3-kpc arm; see][]{2008Dame,Reid19,2021MNRAS.506.2170S}.
Both of the crater-walls traced by EHMCs and the 3-kpc arm 
are naturally located at the similar Galactocentric distance of 
$R_{\rm GC}\sim$~3~kpc, where the high-velocity hot wind is almost stopped
and/or is confined by the cold gas near the gaseous disk.

Due to angular momentum conservation, gas inflows generally accompany with 
gas outflows. The current star formation rate in the CMZ is 
$\lsim 0.1\Msun$~yr$^{-1}$ \citep[e.g.,][]{2009ApJ...702..178Y,2013MNRAS.429..987L},
which will exhaust the CMZ gas in a few $\times 10^{8}$~Myr without other gas supply.
The continuous gas inflow from the inner disk may provide additional 
gas supply for the star formation in the CMZ \citep[e.g., the inflow rate 
at the order of $\sim1-4\ \Msun$~yr$^{-1}$; see][]{2019MNRAS.490.4401A,
2019MNRAS.484.1213S,2020MNRAS.499.4455T,2021ApJ...922...79H}. 

The estimated mass outflow rate based on the CO data is roughly comparable to 
the gas inflow rate from the simulations at the same order of magnitude of 
$\sim2-4\ \Msun$~yr$^{-1}$. Generally, the gas is transported from the inner 
gaseous disk at $R_{\rm GC}\sim$3~kpc 
to the CMZ by inflows, while the angular momentum of gas is taken away by outflows 
from the inner disk. The removed gas in the inner disk may fall back to
the disk \citep[e.g., the fountain models;][]{1976ApJ...205..762S,
1980ApJ...236..577B,1990ApJ...352..506H,2008A&A...484..743S,
2009MNRAS.399.1089M,2015MNRAS.451.4223M,2017ASSL..430..323F}
and/or may be accumulated in the 3-kpc arm and the high-$z$ crater walls.
In this regard, the total molecular gas mass in the CMZ should be roughly 
comparable to that in the 3-kpc arm.

Briefly, the existence of the outflows and the inflows is probably the common feature
in the inner region of the Milky Way (or other barred spiral galaxies).
The dynamical processes indicate that there is a delicate 
balance between gas outflows and inflows toward the inner region of the 
Milky Way. 
Highly variable inflow rate from a recent epoch may lead to the episodic accretion 
onto the CMZ and intermittent activity from Sgr A*
\citep[see, e.g., recent observations in][]{2019Natur.567..347P,2021A&A...646A..66P}.
The following enhanced outflows will then terminate star formation near the GC
and/or restrain nuclear activity due to decreasing the gas supply. 

Finally, Figure~\ref{cartoon} shows a schematic view of the observation results,
i.e., the large-scale $\HI$ voids related to the Fermi bubbles/X-ray bubbles 
in the inner Galaxy, the CO crater walls surrounding the edges of the $\HI$ voids 
above and below the Galactic plane, the thinner gaseous disk within 
$R_{\rm GC}\lsim$3~kpc, the expanding 3-kpc arm at the base of the enhanced EHMCs,
and the entrained MCs with cometary structures pointing away from the Galactic plane.

\section{Summary}
Based on the MWISP CO data and the improved criteria of the DBSCAN algorithm,
we construct high-$z$ MC samples near the tangent points,
in which the distances of the MCs are well determined.
In the region of $l=12^{\circ}$--$26^{\circ}$ and $|b|\lsim5\fdg1$,
a total of 321 high-$z$ MCs (i.e., MCs at $|z|\gsim$110~pc) are identified, of which 47 MCs
lie in the extreme high-$z$ regions (i.e., EHMCs at $|z|\gsim$260~pc).
Besides the weak CO emission and small sizes, these high-$z$ MCs
also display some unusual properties: 

1. The high-$z$ MCs in the $R_{\rm GC}\lsim$3~kpc region are significantly less
than that of the outer region, which is consistent with the
deficient atomic gas and molecular gas in the inner Galactic disk 
of $R_{\rm GC}\lsim$3~kpc.

\begin{figure}
\includegraphics[trim=0mm 0mm 0mm 0mm,scale=0.73,angle=0]{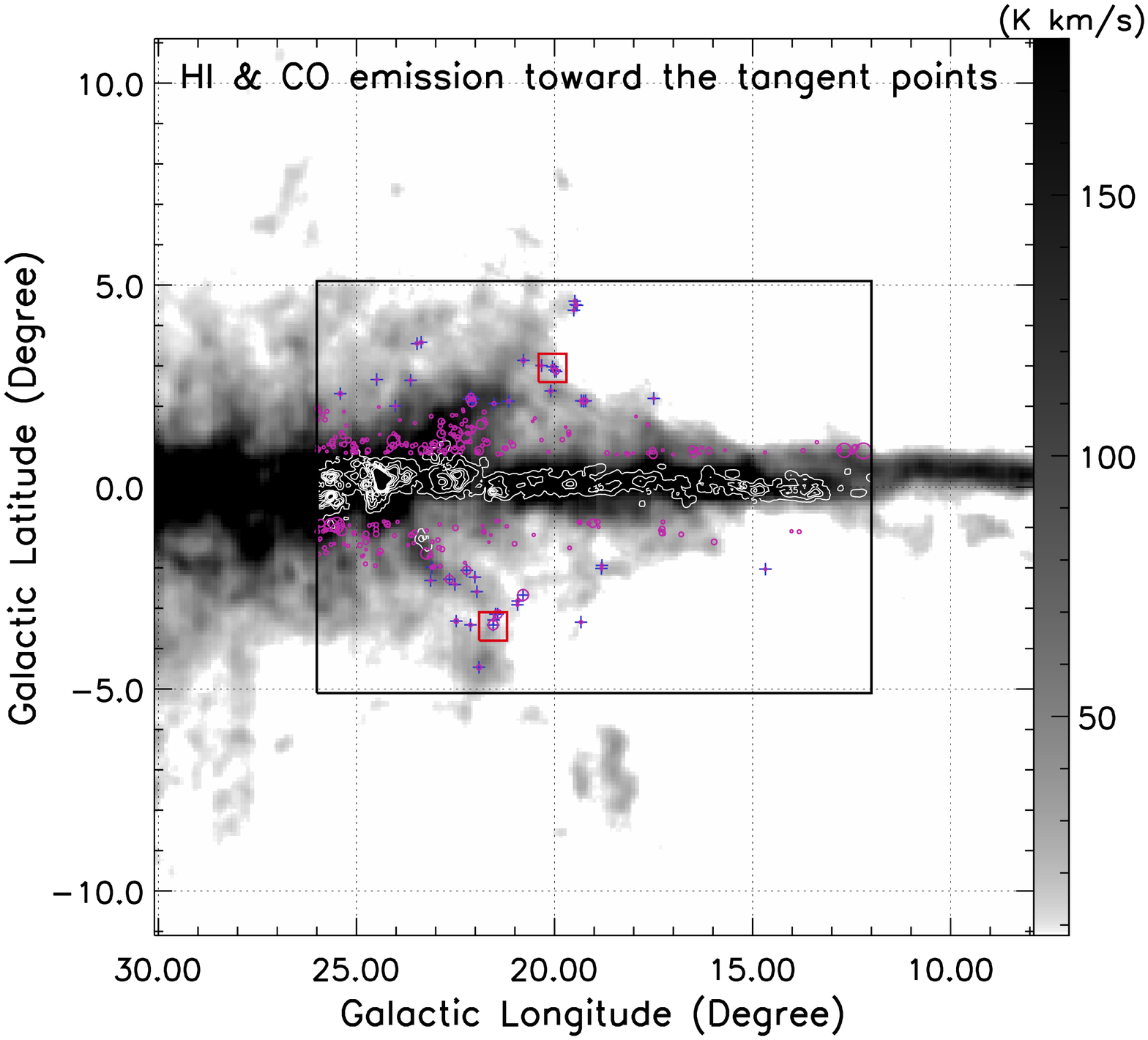}
\caption{
\HI\ (gray) and CO (white contours) emission toward the tangent points.
The atomic and molecular layers are clearly thin for regions of $l\lsim22^{\circ}$. 
The purple circles are 321 high-$z$ MCs ($|z|\gsim$~110~pc) identified from 
the MWISP data in the region of $l=12^{\circ}$--$26^{\circ}$ and $|b|\lsim5\fdg1$
(the black rectangle).
Note that the circle's size is not the true angular size of the MCs,
but it is proportional to the effective radius of the clouds
(i.e., $d\times\sqrt{(\theta_{\rm MC}^2-\theta_{\rm beam}^2)/\pi}$, where $\theta_{\rm MC}$
and $\theta_{\rm beam}$, in units of arcmin, are the angular size of the CO emission
and the beam size, respectively). 
The blue plus signs indicate the positions of the 47 identified EHMCs 
(see Section 3.3 and Table~1).
The red boxes indicate two zoom-in regions shown in Figure~3. 
\label{tangent}}
\end{figure}

2. The EHMCs (i.e., IDs 05--31 in Table~1) are mainly concentrated in 
narrow regions of [$l\sim$19\fdg1 to 20\fdg5, 
$b\sim$2\fdg0 to 5\fdg1] and [$l\sim$20\fdg5 to 22\fdg1, $b\sim-$2\fdg0 to $-5$\fdg1].
Some EHMCs are even located at $|z|\gsim$~600~pc far above and below the Galactic plane.
The EHMC concentrations, together with other high-$z$ MCs at 
$l\lsim18^{\circ}$, constitute molecular crater-walls with a 
measured thickness of $\sim$~220~pc.
The molecular crater wall structures lie along the edges of the $\HI$ voids 
(Figure~\ref{tangent}) that are associated with the Milky Way nuclear wind
\citep[e.g.,][]{2016ApJ...826..215L,2021MNRAS.506.2170S}.

3. Some large high-$z$ MCs, which lie in the crater walls, 
display intriguing elongated head-to-tail structures pointing away from the Galactic disk,
favoring the scenario of the entrained molecular gas moving with the multiphase outflows.
Especially, the $\sim20$~pc long tail of the EHMC G021.548$-$03.414 (Figure~\ref{twosamples}) 
is physically associated with a filamentary structure of the IR dust emission,
which is exactly located in the enhanced $\HI$ ridge.
The observed velocity gradient of the EHMC (Figure~\ref{pv}), together with its
cometary head toward the Galactic plane, shows that the cold molecular gas
is indeed entrained by the multiphase outflows from the Galactic plane 
to the high-$z$ regions. 

Based on the above results, we suggest that the powerful nuclear wind of 
the Milky Way has a profound impact on the large-scale distribution of 
the gaseous disk (Figure~\ref{cartoon}). 
The \HI\ voids above and below the Galactic plane, the CO crater walls
at the edges of the \HI\ voids, and the expanding 3-kpc arms at
the base of the molecular crater walls are probably the natural result
of the intermittent nuclear activity of the Milky Way in the recent 3--6~Myr.

The cometary MCs lying in the crater walls show that the nuclear wind 
removes gas from the inner Galaxy to the high-$z$ regions.
The cold gas in the crater-wall structures at $R_{\rm GC}\sim$3~kpc plays 
a crucial role confining the Milky Way nuclear wind. 
The cool outflows may be an important mass reservoir 
for supplying halo material that has been pushed up from the interface 
between the nuclear wind and the gaseous disk within $R_{\rm GC}\lsim$3~kpc. 
In this scenario, some interesting estimates can be summarized as follows: 

1. The gas-rich environment increases the survival ability of the EHMCs 
in the crater walls with the local high density of $\sim$1--10~cm$^{-3}$
for \HI\ clouds and $\sim 10^2-10^3$~cm$^{-3}$ for CO clouds. 
The estimated lifetime of EHMCs is several Myr for clouds on the parsec scales, 
which is comparable to the dynamical time of the cool gas flows according to 
the observed velocity gradient of the CO gas 
(i.e., $\sim0.5-1\km\ps$pc$^{-1}$ after the projection correction
with a small inclination angle of $i\sim20^{\circ}-10^{\circ}$).
Basically, the observed EHMCs in the walls will be destroyed in future several Myr. 
The MCs thus cannot move too far away from the Galactic plane 
(e.g., $|z|\gsim$~1~kpc) before the molecular gas becomes 
the neutral atomic gas and/or ionized gas.

2. The velocity of the cool outflows is estimated to be $\sim140-330\km\ps$ 
assuming that the gas is launched from the boundary of the thin CO disk
\citep[i.e., from $|z|=3\times\sigma_z \sim$110-120~pc;][]{2021ApJ...910..131S}.
The hypothesis is supported by the observed cometary high-$z$ MCs pointing away 
from the Galactic plane at locations of $|z|\sim$110--130~pc 
(top panels of Figure~\ref{lb4s}) and the large-scale velocity gradient of 
the coherent EHMCs at $|z|\sim$400--450~pc far from the plane (see Figure~\ref{tri}).

\begin{figure}
\includegraphics[trim=0mm 0mm 0mm 20mm,scale=0.8,angle=90]{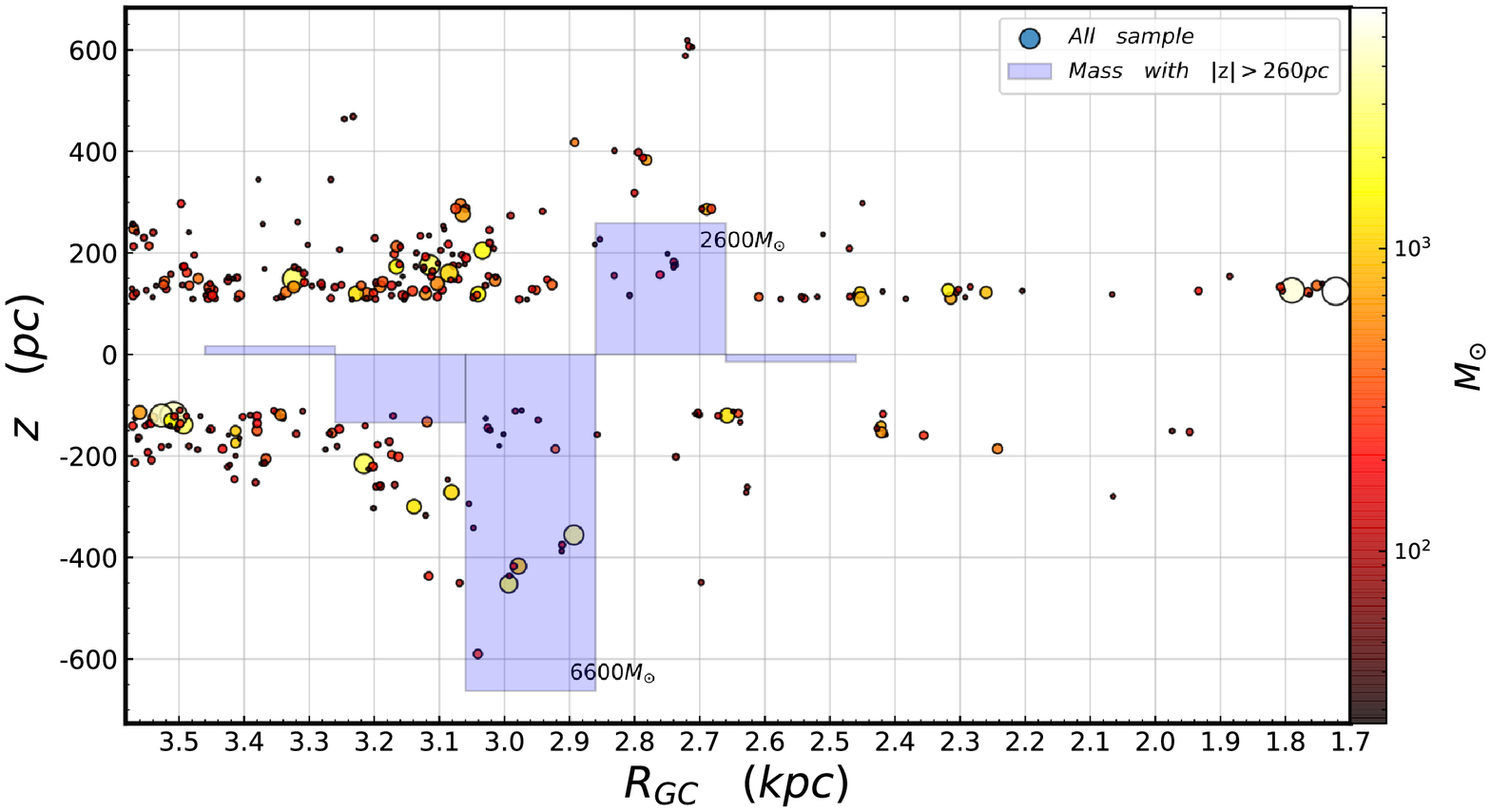}
\caption{
321 high-$z$ MCs in the $R_{\rm GC}$--$z$ map. The size of the filled circles
is proportional to the effective radius of the MCs, while the color 
represents the mass of the clouds. The histogram shows the mass distribution
of the MCs at $|z|\gsim$~260~pc (about $3\times$FWHM of the thin CO disk).
\label{lb}}
\end{figure}

\begin{figure}
\gridline{\fig{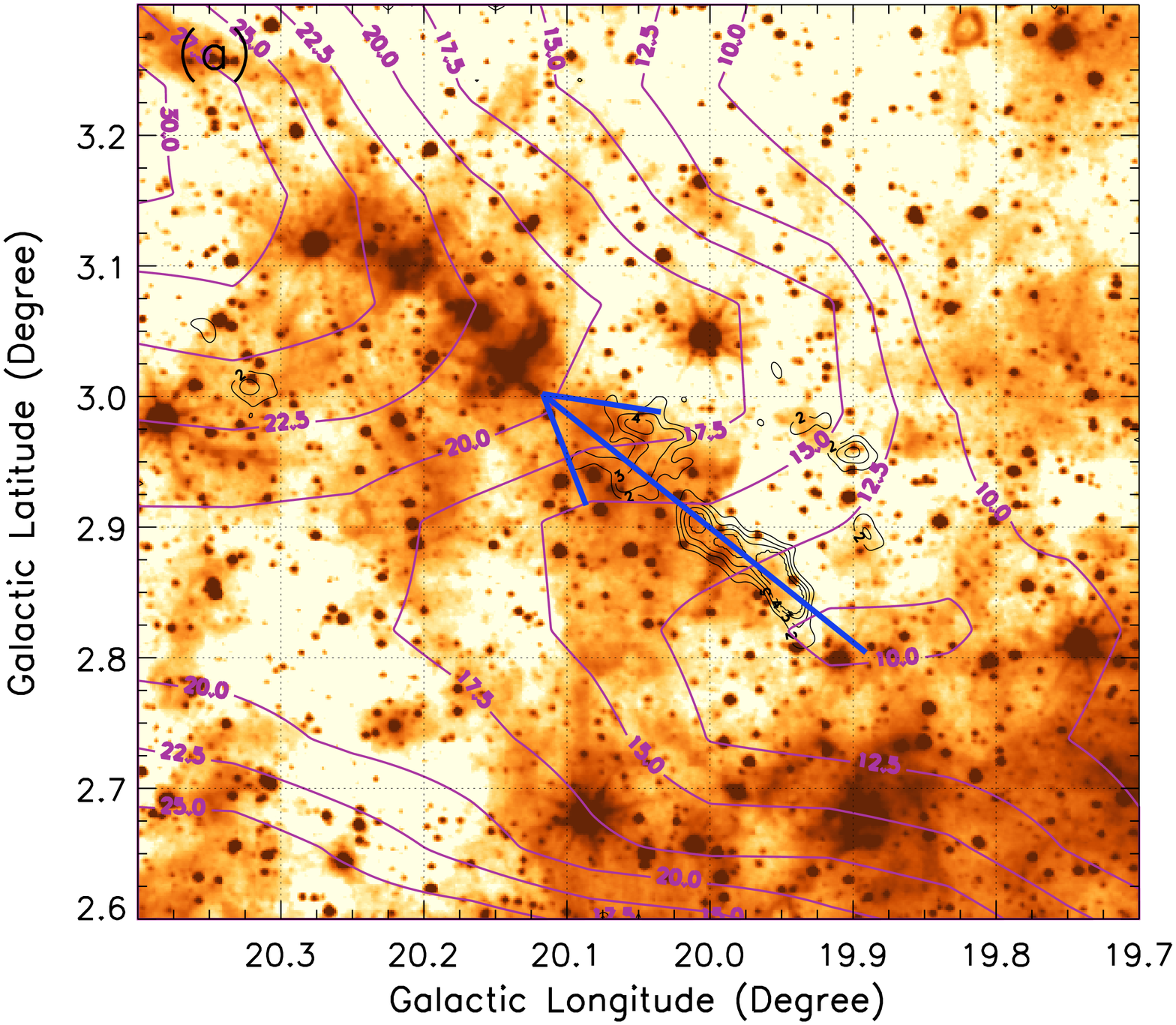}{0.5\textwidth}{}}
\gridline{\fig{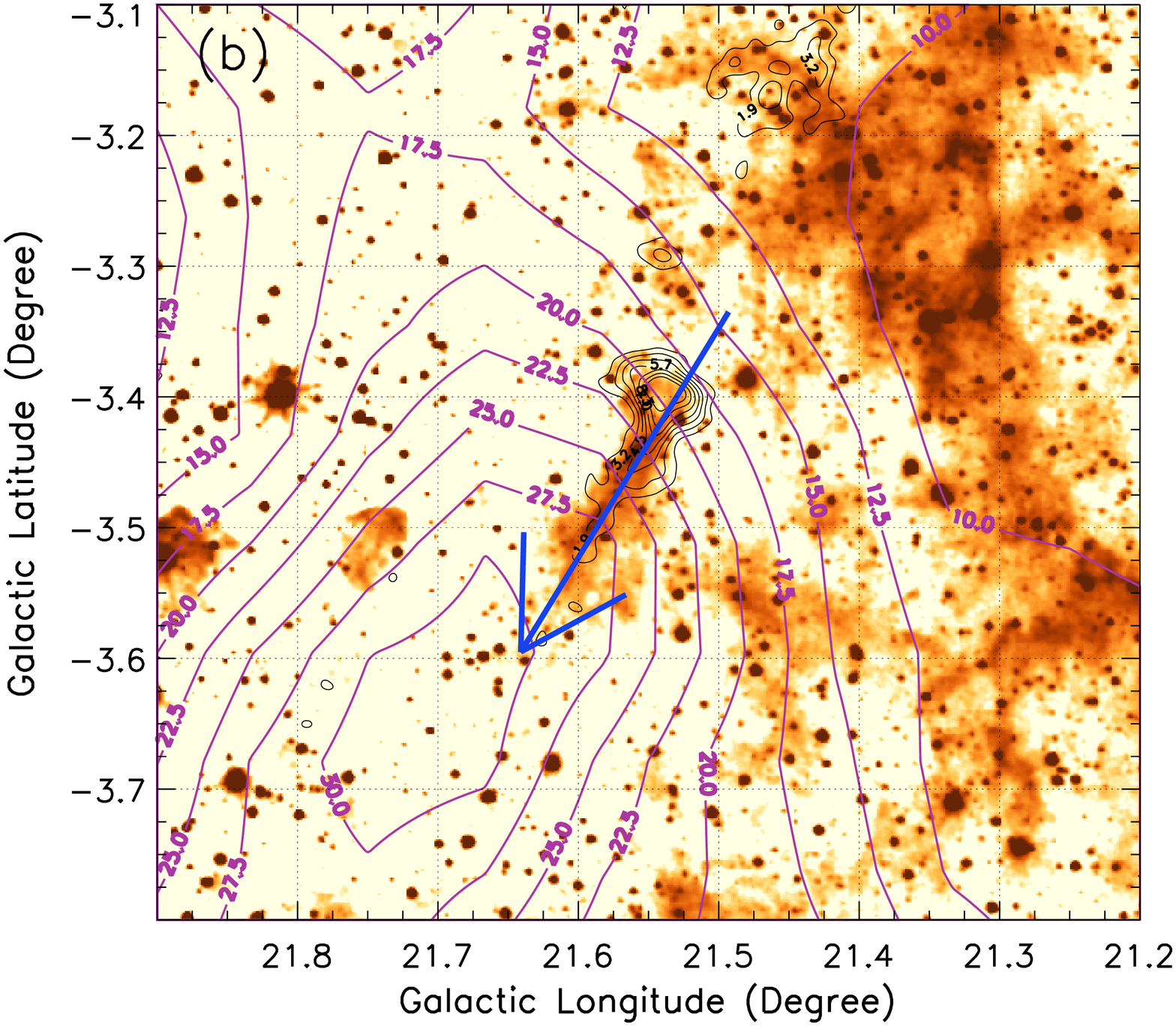}{0.5\textwidth}{}}
\caption{
Multiwavelength views of two EHMCs toward the tangent points.
The images display WISE 12~$\mu$m emission overlaid with 
the \twCO\ integrated emission (black contours) and 
\HI\ emission (purple contours) for the cometary clouds
above (panel a for EHMC G019.957$+$02.863 in the velocity interval 
of 124--136~$\km\ps$) and below (panel b for EHMC G021.548$-$03.414 
in the velocity interval of 114--125~$\km\ps$) the Galactic plane 
(see the red boxes in Figure~\ref{tangent}).
The blue arrows indicate the direction of the cometary structures
from the head (toward the Galactic plane) to the tail.
The PV diagrams along the arrows are shown in Figure~\ref{pv}.
\label{twosamples}}
\end{figure}

\begin{figure}
\gridline{\fig{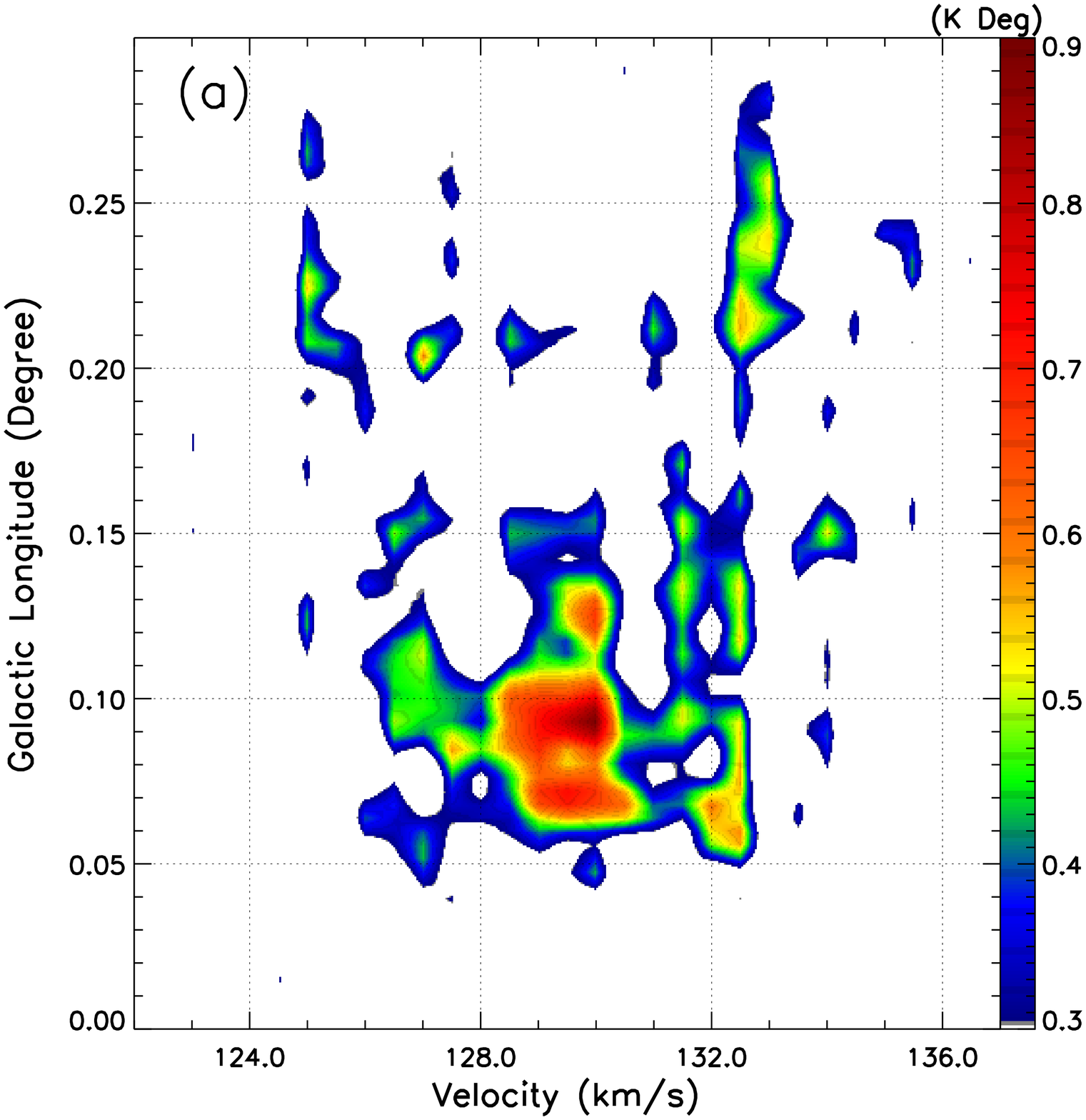}{0.4\textwidth}{}}
\gridline{\fig{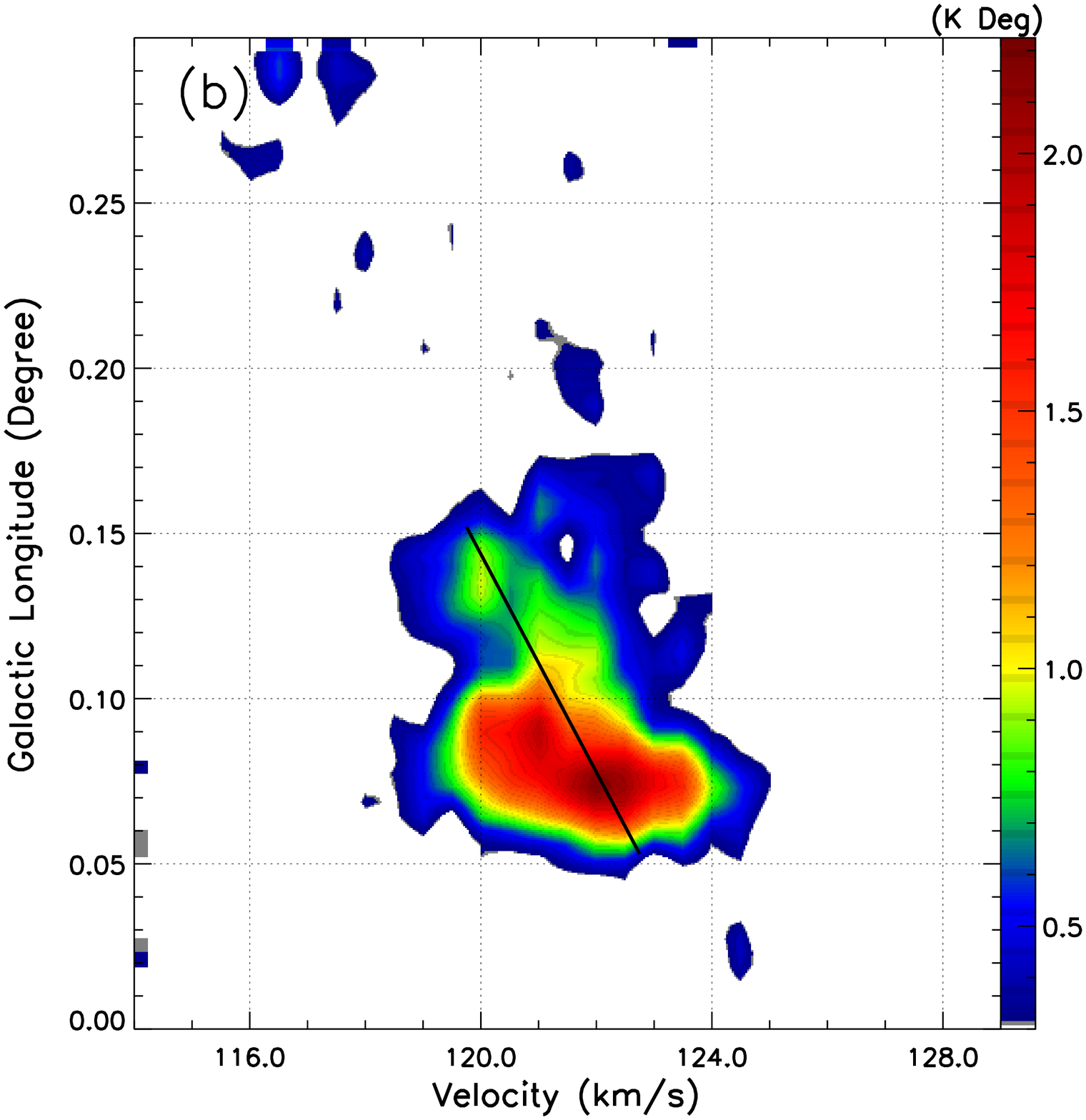}{0.4\textwidth}{}}
\caption{
PV diagrams of the \twCO\ emission
of two EHMCs shown in Figure~\ref{twosamples},
i.e., panel (a) from ($l=$19\fdg89, $b=+$2\fdg80) to ($l=$20\fdg12, $b=+$3\fdg00)
for EHMC G019.957$+$02.863,
and panel (b) from ($l=$21\fdg49, $b=-$3\fdg34) to ($l=$21\fdg64, $b=-$3\fdg60)
for EHMC G021.548$-$03.414.
The two slices have a length of 0\fdg3 and a width of 3\farcm5.
The black line in panel (b) displays the velocity gradient of 
$\sim -0.23\km\ps$pc$^{-1}$ at the tangent distance of $\sim 7.6$~kpc.
\label{pv}}
\end{figure}

\begin{figure}
\includegraphics[trim=20mm 0mm 0mm 40mm,scale=0.8,angle=0]{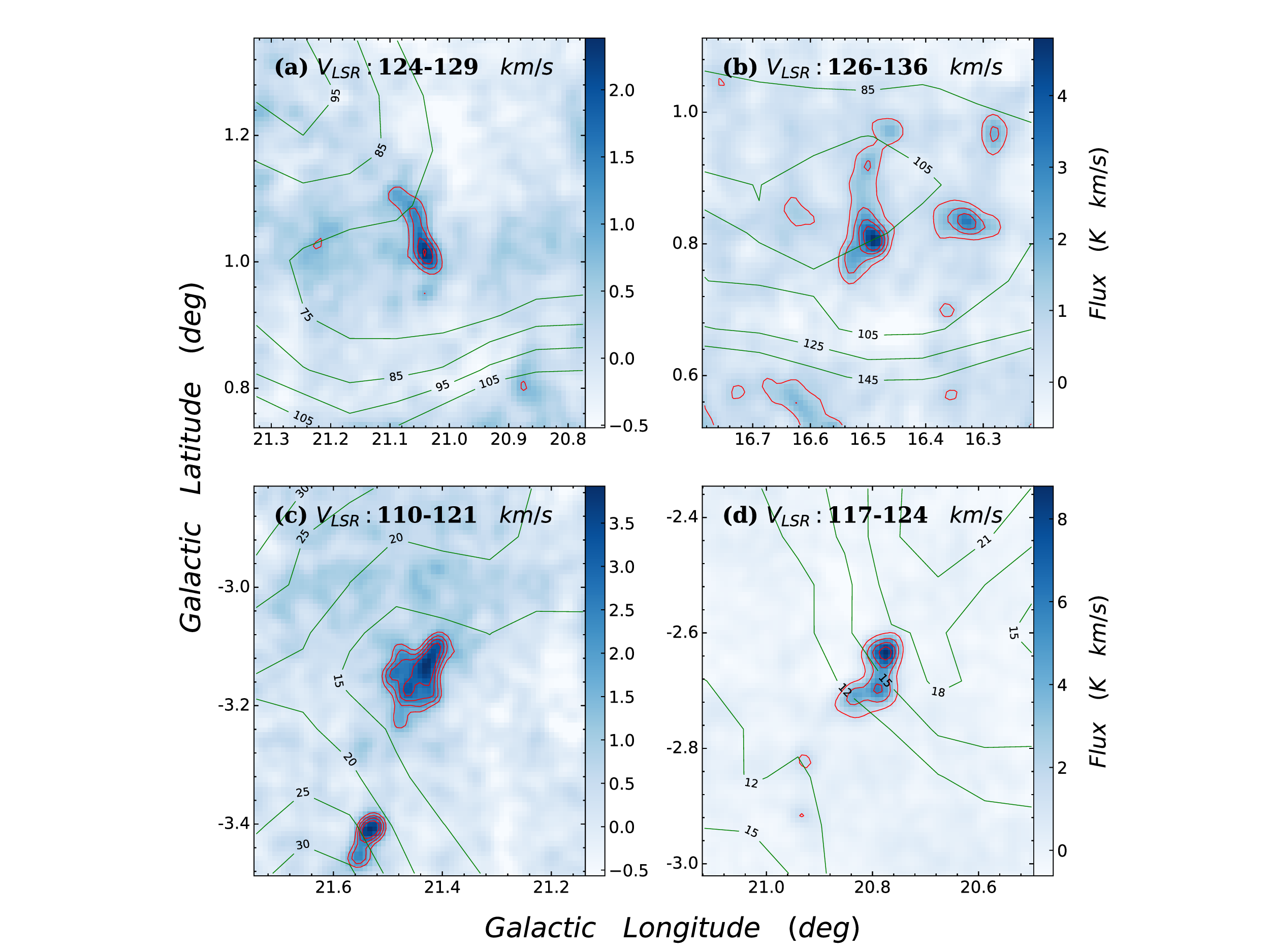}
\caption{
CO emission of four elongated high-$z$ MCs, overlaid with the 
\HI\ emission near the tangent points (green contours in units of K~$\km\ps$). 
The integrated velocity ranges of the CO emission are labeled in the upper left 
corner of each panel. The CO red contours are 
(a) 0.8, 1.3, 1.8, 2.3~K~$\km\ps$; (b) 1.0, 1.8, 2.6, 3.4~K~$\km\ps$;
(c) 1.5, 2.0, 2.5, 3.0~K~$\km\ps$; and (d) 1.5, 3.5, 5.5, 7.5~K~$\km\ps$, respectively.
\label{lb4s}}
\end{figure}

\begin{figure}
\includegraphics[trim=0mm 0mm 0mm 0mm,scale=0.7,angle=90]{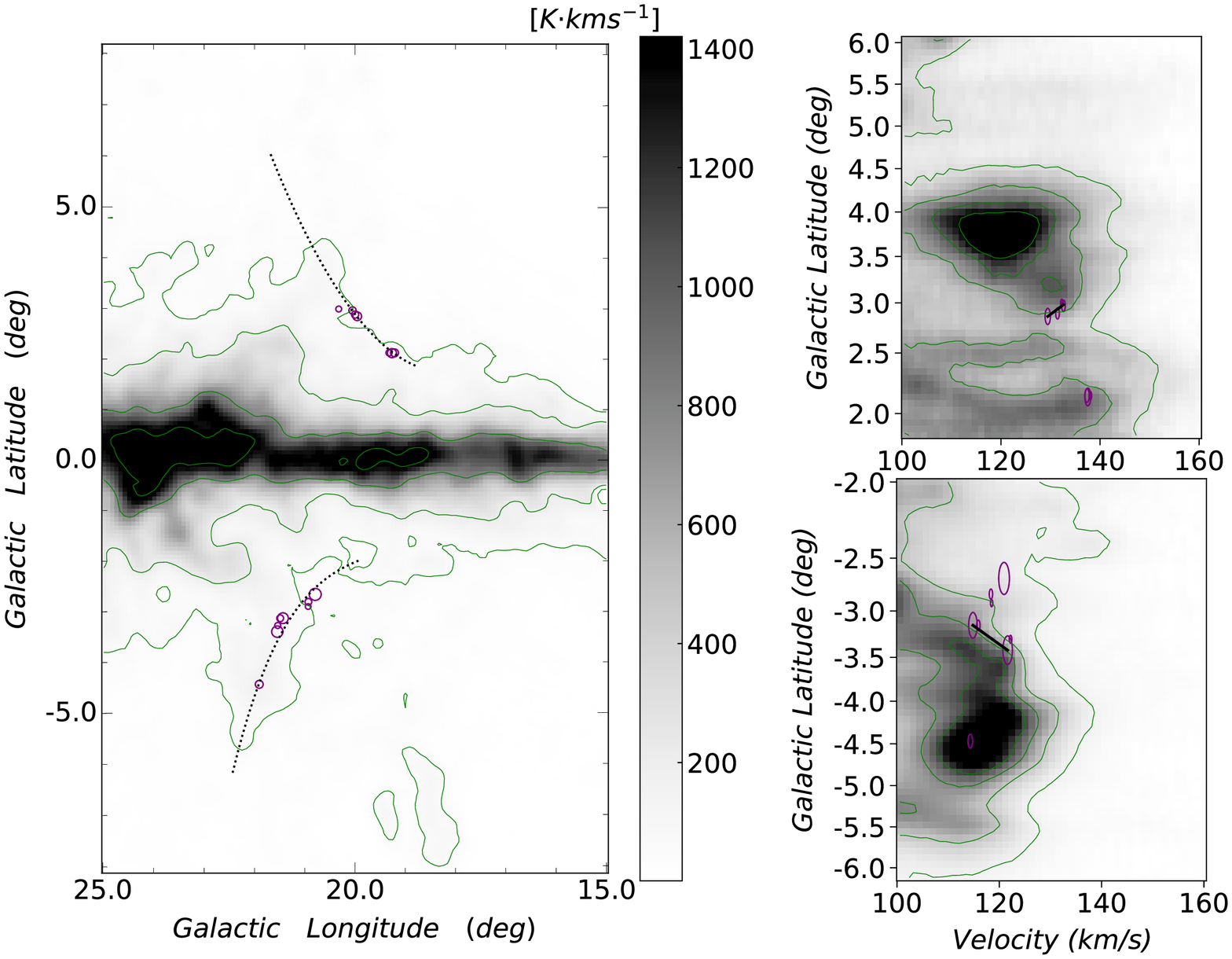}
\caption{
Left panel: \HI\ intensity map overlaid with the EHMCs (purple circles) 
along the crater walls (black dashed lines).
The green contours indicate the \HI\ intensity of 50, 200, 800, and 1500 K~$\km\ps$
toward the tangent points.
Top right panel: the PV diagram along the crater wall (with a width of 35$'$) 
above the Galactic plane. The green contours are 0.6, 1.2, 1.8, and 2.4~K.
The purple ellipses are the same EHMCs in the left panel 
(i.e., ID 5, 6, 7, 13, 14, 15, and 17 in Table~1).
The black solid line indicates the velocity gradient of $\sim 0.15\km\ps$pc$^{-1}$
for the coherent EHMCs.
Bottom right panel: same as the top right panel, but for the crater wall
below the plane. The green contours are 0.6, 1.8, 3.0, and 4.2~K.
The purple ellipses are from Table~1
(ID 19, 20, 21, 23, 24, 26, 27, and 28). The velocity gradient
is $\sim 0.16\km\ps$pc$^{-1}$.
Note that the EHMC's size is not the true angular size of the detected CO emission,
but it is proportional to the effective radius of the clouds.
\label{tri}}
\end{figure}

\begin{figure}
\includegraphics[trim=0mm 0mm 0mm 0mm,scale=0.6,angle=90]{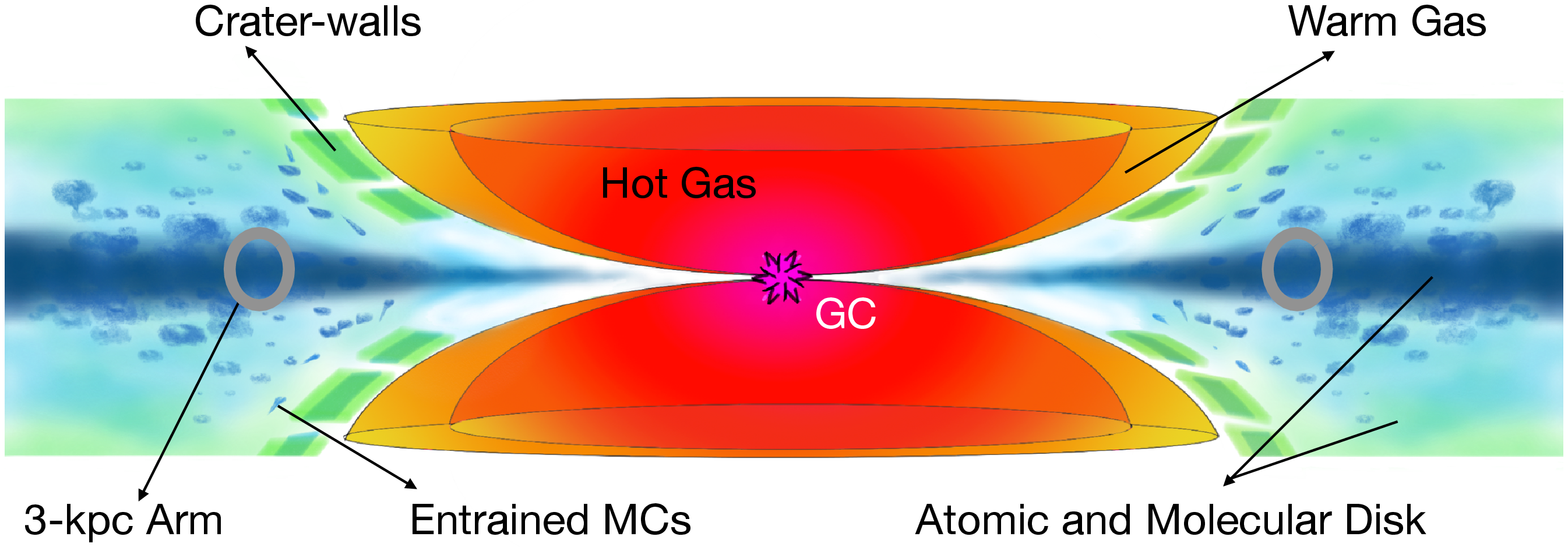}
\caption{
Schematic diagram of the relationship between the Milky Way nuclear wind
and the gaseous disk (green is for the atomic gas, and blue is for the 
molecular gas). 
Note that the size of the MCs in the diagram is not exactly to scale 
because we aim to highlight the head-to-tail MCs pointing away from the 
Galactic plane.  
The same applies for the sizes of the crater walls.
\label{cartoon}}
\end{figure}

3. The molecular gas in the EHMC concentrations of 
$\sim2\times220$~pc$\times410$~pc 
has a total mass of $\gsim1\times10^{4}\ \Msun$ (Figure~\ref{lb}).
If the extreme high-$z$ MCs are more broadly distributed in
the whole regions of $R_{\rm GC}\sim$~3~kpc, $|z|\sim260$--670~pc, and
the wall's thickness of $\sim$~220~pc,
the total molecular mass is estimated to be $\gsim1\times10^{6}\ \Msun$,
which is comparable to the total \HI\ mass in the Fermi bubbles
\citep[][]{2018ApJ...855...33D,2020ApJ...888...51L}.

4. Assuming a Gaussian distribution for the thick CO disk \citep[i.e., $\sigma_z=$120~pc 
and the thickness FWHM=2.355$\sigma_z$ in][]{2021ApJ...910..131S},
a significant amount of molecular gas (e.g., the order of $10^{7}\ \Msun$) 
may accumulate at the low latitudes of the gaseous disk of $|z|\sim$110--260~pc.
The mass-loading rate of the cool outflows at $R_{\rm GC}\sim$~3~kpc
(i.e., outflows to the crater-wall structures) is roughly comparable to the mass inflow rate 
(i.e., inflows to the CMZ) at the same order of $\sim2-4\ \Msun$~yr$^{-1}$.

5. The thinner gas disk within $R_{\rm GC}$\lsim3~kpc may be the joint result of 
(1) inflows from the inner gaseous disk to the CMZ and 
(2) outflows from the gaseous disk to the 3-kpc arm and the high-$z$ region.
The 3-kpc arm at the base of the EHMC concentration, together with the thinner 
gaseous disk within $R_{\rm GC}\lsim$3~kpc, can be naturally
explained by the interaction between the Milky Way nuclear
wind and the Galactic gaseous disk.

Considering the large uncertainties in the discussions, 
the above estimates should be used with caution.
Nevertheless, we think that the results are useful for further studies. 
For example, the multiwavelength observations (e.g., radio continuum, 
millimeter and submillimeter molecular line emission, optical/near-IR 
emission lines, and UV absorption) are advocated to
investigate the physical properties of the cometary high-$z$ MCs. The large-scale
surveys with high sensitivity and high resolution, together with the improved simulations, 
are also very helpful to reveal the nature of the Galactic nuclear winds/outflows.

\acknowledgments
This research made use of the data from the Milky Way Imaging Scroll Painting 
(MWISP) project, which is a multiline survey in \twCO/\thCO/C$^{18}$O along the 
northern Galactic plane with the PMO 13.7m telescope. We are grateful to all the members 
of the MWISP working group, particularly the staff members at the PMO 13.7m telescope, 
for their long-term support. MWISP was sponsored by the National Key R\&D Program of 
China with grant 2017YFA0402700 and the CAS Key Research Program of Frontier Sciences 
with grant QYZDJ-SSW-SLH047.
We acknowledge support from the National Natural Science Foundation of China 
through grants 12173090 and 12041305. 
X.C. acknowledges support by the CAS International Cooperation Program
(grant No. 114332KYSB20190009).
We also thank the anonymous referee for many useful and
constructive comments that largely improved the quality of the paper.

The work makes use of publicly released data from the HI4PI survey, which combines 
the EBHIS in the Northern hemisphere with the GASS in the Southern Hemisphere.
The Parkes Radio Telescope is part of the Australia Telescope National Facility, 
which is funded by the Australian Government for operation as a National Facility 
managed by CSIRO. The EBHIS data are based on observations performed with the 
100 m telescope of the MPIfR at Effelsberg. EBHIS was funded by the Deutsche 
Forschungsgemeinschaft (DFG) under the grants KE757/7-1 to 7-3.
This publication makes use of data products from the Wide-field Infrared Survey 
Explorer, which is a joint project of the University of California, Los Angeles, 
and the Jet Propulsion Laboratory/California Institute of Technology,
funded by the National Aeronautics and Space Administration.

\facility{PMO 13.7m}
\software{GILDAS/CLASS \citep{2005sf2a.conf..721P}} 

\bibliographystyle{aasjournal}
\bibliography{references}

\clearpage

\clearpage

\clearpage

\clearpage

\clearpage

\clearpage

\clearpage

\begin{deluxetable}{ccccccccccc}
\tabletypesize{\scriptsize}
\tablecaption{Parameters of 47 EHMCs based on the MWISP \twCO ($J$=1--0) Emission}
\tablehead{
\colhead{\begin{tabular}{c}
ID  \\
    \\
(1)  \\
\end{tabular}} &
\colhead{\begin{tabular}{c}
$l$               \\
(deg)      \\
(2)  \\
\end{tabular}} &
\colhead{\begin{tabular}{c}
$b$               \\
(deg)      \\
(3)  \\
\end{tabular}} &
\colhead{\begin{tabular}{c}
$v_{\rm LSR}$      \\
($\km\ps$)         \\
(4)  \\
\end{tabular}} &
\colhead{\begin{tabular}{c}
$\sigma_v$     \\
($\km\ps$)                  \\
(5)  \\
\end{tabular}} &
\colhead{\begin{tabular}{c}
$T_{\rm peak}$   \\
(K)              \\
(6)  \\
\end{tabular}} &
\colhead{\begin{tabular}{c}
Radius$^{\mathrm {a}}$          \\
(pc)     \\
(7)  \\
\end{tabular}} &
\colhead{\begin{tabular}{c}
Distance$^{\mathrm {b}}$            \\
(kpc)     \\
(8)  \\
\end{tabular}} &
\colhead{\begin{tabular}{c}
$z$ Height    \\
(pc)     \\
(9)  \\
\end{tabular}} &
\colhead{\begin{tabular}{c}
Mass           \\
($\Msun$)     \\
(10)  \\
\end{tabular}} &
\colhead{\begin{tabular}{c}
$\alpha^{\mathrm {c}}$     \\
     \\
(11)  \\
\end{tabular}}
}
\startdata
01  &  14.675  &  $-$2.031  &   142.66 	  &  0.52   &  2.03   &   1.7 	  &     7.9   & 	$-$280    &  	71 	  &  7.6 	  \\
02  &  17.495  &  2.194   &  	141.26 	  &  0.67   &  1.72   &   1.7 	  &  	7.8   & 	298 	  &  	76 	  &  11.6 	  \\
03  &  18.803  &  $-$1.939  &   131.30 	  &  0.61   &  1.69   &   1.9 	  &  	7.7   & 	$-$261    &  	70 	  &  11.7 	  \\
04  &  18.815  &  $-$2.017  &   128.91 	  &  0.83   &  1.40   &   1.8 	  &  	7.7   & 	$-$272    &  	75 	  &  19.1 	  \\
05  &  19.214  &  2.135   &  	137.37 	  &  1.24   &  1.62   &   3.5 	  &  	7.7   & 	287 	  &  	322       &  19.4 	  \\
06  &  19.267  &  2.129   &  	137.53 	  &  2.35   &  1.93   &   4.6 	  &  	7.7   & 	286 	  &  	789       &  37.6 	  \\
07  &  19.324  &  2.135   &  	138.05 	  &  1.04   &  1.85   &   2.1 	  &  	7.7   & 	287 	  &  	120       &  22.3 	  \\
08  &  19.332  &  $-$3.343  &   129.38 	  &  0.60   &  1.53   &   2.1 	  &  	7.7   & 	$-$449    &  	74 	  &  12.2 	  \\
09  &  19.430  &  4.502   &  	129.72 	  &  0.97   &  1.32   &   1.4 	  &  	7.7   & 	605 	  &  	53 	  &  28.9 	  \\
10  &  19.468  &  4.515   &  	128.17 	  &  1.25   &  2.05   &   2.6 	  &  	7.7   & 	607 	  &  	153       &  30.8 	  \\
11  &  19.489  &  4.605   &  	129.97 	  &  0.81   &  1.53   &   1.8 	  &  	7.7   & 	619 	  &  	58 	  &  23.7 	  \\
12  &  19.512  &  4.378   &  	137.76 	  &  0.73   &  2.08   &   1.8 	  &  	7.7   & 	588 	  &  	104       &  10.7 	  \\
13  &  19.957  &  2.863   &  	129.44 	  &  1.63   &  1.88   &   4.2 	  &  	7.7   & 	383 	  &  	620       &  21.0 	  \\
14  &  19.999  &  2.897   &  	131.42 	  &  0.91   &  1.52   &   2.9 	  &  	7.7   & 	388 	  &  	139       &  19.9 	  \\
15  &  20.051  &  2.978   &  	132.61 	  &  0.64   &  1.30   &   2.9 	  &  	7.7   & 	398 	  &  	107       &  13.1 	  \\
16  &  20.095  &  2.380   &  	124.45 	  &  1.07   &  1.67   &   2.6 	  &  	7.7   & 	318 	  &  	133       &  26.4 	  \\
17  &  20.326  &  3.007   &  	132.28 	  &  0.71   &  1.64   &   1.9 	  &  	7.6   & 	402 	  &  	65 	  &  17.1 	  \\
18  &  20.785  &  3.138   &  	115.74 	  &  1.58   &  1.93   &   3.1	  &  	7.6   & 	418 	  &  	466       &  19.4 	  \\
19  &  20.794  &  $-$2.672  &   120.93 	  &  1.27   &  4.61/1.26$^{\mathrm {d}}$   &   8.1 	  &  	7.6   & 	$-$356    &  	3118      &  4.9         \\
20  &  20.930  &  $-$2.821  &   118.36 	  &  0.67   &  1.36   &   2.6 	  &  	7.6   & 	$-$375    &  	128       &  10.2 	  \\
21  &  20.934  &  $-$2.914  &  	118.50 	  &  0.65   &  1.82   &   1.8 	  &  	7.6   & 	$-$388    &  	75 	  &  11.7 	  \\
22  &  21.156  &  2.125   &  	114.33 	  &  0.90   &  1.61   &   2.2 	  &  	7.6   & 	282 	  &  	103       &  20.2 	  \\
23  &  21.436  &  $-$3.146  &  	114.85 	  &  1.84   &  1.78   &   6.5 	  &     7.6   & 	$-$417    &  	1012      &  25.4        \\
24  &  21.491  &  $-$3.147  &  	115.91 	  &  1.39   &  1.32   &   2.6 	  &  	7.6   & 	$-$417    &  	170       &  34.3 	  \\
25  &  21.526  &  2.067   &  	137.91 	  &  0.95   &  1.17   &   2.5 	  &  	7.6   & 	274 	  &  	101       &  26.3 	  \\
26  &  21.540  &  $-$3.291  &   122.20 	  &  0.98   &  1.43   &   2.0 	  &     7.6   & 	$-$436    &  	108       &  20.5 	  \\
27  &  21.548  &  $-$3.414  &  	121.63 	  &  1.39   &  4.36/1.14$^{\mathrm {d}}$   &   7.2 	  &  	7.6   & 	$-$452    &  	2340      &   7.0        \\
28  &  21.908  &  $-$4.462  &  	114.33 	  &  0.92   &  2.72   &   3.5 	  &  	7.6   & 	$-$590    &  	281       &  12.3 	  \\
29  &  21.959  &  $-$2.589  &  	125.82 	  &  0.77   &  1.59   &   1.9 	  &  	7.6   & 	$-$342    &  	73 	  &  17.6 	  \\
30  &  22.011  &  $-$2.231  &   114.55 	  &  0.71   &  1.45   &   1.8 	  &  	7.6   & 	$-$294    &  	71 	  &  14.4 	  \\
31  &  22.050  &  2.185   &  	128.01 	  &  0.45   &  1.20   &   3.0 	  &     7.6   & 	288 	  &  	80 	  &   8.9 	  \\
32  &  22.085  &  2.095   &  	130.88 	  &  0.86   &  2.34   &   6.2 	  &  	7.6   & 	276 	  &  	685       &  	7.7 	  \\
33  &  22.110  &  2.242   &  	127.66 	  &  1.30   &  2.03   &   4.5 	  &  	7.6   & 	296 	  &  	443       &  20.1 	  \\
34  &  22.121  &  $-$3.413  &  	116.77 	  &  0.56   &  1.49   &   2.5 	  &     7.6   & 	$-$450    &  	102       &  	9.1 	  \\
35  &  22.166  &  2.181   &  	131.57 	  &  0.71   &  2.15   &   4.0 	  &  	7.5   & 	288 	  &  	316       &  	7.5 	  \\
36  &  22.216  &  $-$2.061  &  	113.76 	  &  1.64   &  2.18   &   6.1 	  &     7.5   & 	$-$272    &  	1114      &  17.0         \\
37  &  22.481  &  $-$3.318  &   115.78 	  &  0.69   &  2.12   &   3.2 	  &  	7.5   & 	$-$437    &  	231       &  	7.8 	  \\
38  &  22.515  &  $-$2.412  &   111.25 	  &  0.68   &  1.18   &   1.9 	  &  	7.5   & 	$-$317    &  	44 	  &  22.7 	  \\
39  &  22.654  &  $-$2.283  &  	111.06 	  &  1.75   &  2.32   &   5.8 	  &     7.5   & 	$-$300    &  	1262      &  16.4         \\
40  &  23.098  &  $-$1.990  &  	125.68 	  &  1.30   &  1.86   &   2.4 	  &  	7.5   & 	$-$261    &  	113       &  42.1 	  \\
41  &  23.128  &  $-$2.314  &  	121.09 	  &  0.47   &  1.14   &   1.8 	  &     7.5   &  	$-$303    & 	38 	  &  11.7 	  \\
42  &  23.367  &  3.585   &  	117.59 	  &  0.58   &  1.24   &   2.2 	  &  	7.5   & 	469 	  &  	68 	  &  12.5 	  \\
43  &  23.470  &  3.550   &  	115.42 	  &  0.44   &  1.54   &   1.9 	  &     7.5   & 	464 	  &  	50 	  &   8.2 	  \\
44  &  23.632  &  2.642   &  	111.06 	  &  0.75   &  1.00   &   2.1 	  &  	7.5   & 	345 	  &  	54 	  &  25.3 	  \\
45  &  24.021  &  2.006   &  	121.30 	  &  0.55   &  1.51   &   1.9 	  &     7.4   & 	261 	  &  	66 	  &   9.9 	  \\
46  &  24.488  &  2.661   &  	109.69 	  &  0.36   &  1.67   &   1.5 	  &     7.4   & 	345 	  &  	43 	  &   5.3 	  \\
47  &  25.408  &  2.311   &  	108.45 	  &  1.31   &  1.57   &   2.8 	  &  	7.4   & 	297	  &  	193       &  29.1 	  \\
\hline
\enddata
\tablecomments{
$^{\mathrm {a}}$ The effective radius of the clouds.
$^{\mathrm {b}}$ The error of the estimated distance is about 2\%--20\%
assuming that the MCs are located near tangent points
with a velocity uncertainty of $\sim5\km\ps$
along the line of sight \citep[see, e.g., the A5 model in][]{Reid19}.
$^{\mathrm {c}}$ The MC's virial parameter estimated from the definition of
$\alpha=\frac{5\sigma_{v}^2R}{GM_{\rm Xfactor}}$ (see Section 3.3).
$^{\mathrm {d}}$ Detected \thCO\ emission associated with \twCO\ emission.
The \thCO\ emission is very weak, i.e., $T_{\rm peak13}\sim$~1~K.
}
\end{deluxetable}

\end{document}